\documentclass{aastex}
\shorttitle{Herschel/PACS Observations of protoplanetary disks in Taurus/ Auriga}
\shortauthors{Howard et al.}
\begin{document}

\def\Msun{\,{\rm M$_{\odot}$}}
\def\Lsun{\,{\rm L$_{\odot}$}}
\def\psec{$^s\mskip-7.6mu.\,$}
\def\ptmin{$'\mskip-7.6mu.\,$}

\title{{\it Herschel}/PACS Survey of protoplanetary disks in Taurus/Auriga -- Observations of  [\ion{O}{1}] and [\ion{C}{2}], and far infrared continuum}

\slugcomment{Herschel is an ESA space observatory with science instruments provided
by European-led Principal Investigator consortia and with important
participation from NASA.}
\author{Christian D. Howard\altaffilmark{1,2}, G{\"o}ran Sandell\altaffilmark{1}, William D.Vacca\altaffilmark{1},  
Gaspard Duch\^ene\altaffilmark{3,4}, Geoffrey Mathews\altaffilmark{5,6}, Jean-Charles Augereau\altaffilmark{4}, 
David Barrado\altaffilmark{7,8}, William R. F.Dent\altaffilmark{9},  Carlos Eiroa\altaffilmark{10}, Carol Grady\altaffilmark{11,12,13}, 
Inga Kamp\altaffilmark{14}, Gwendolyn Meeus\altaffilmark{10}, 
Francois M{\'e}nard\altaffilmark{4,15}, Christophe Pinte\altaffilmark{4}, Linda Podio\altaffilmark{4}, 
Pablo Riviere-Marichalar\altaffilmark{7}, Aki Roberge\altaffilmark{11},  Wing-Fai Thi\altaffilmark{4},
Silvia Vicente\altaffilmark{14}, Jonathan P. Williams\altaffilmark{6}}

\altaffiltext{1}{SOFIA-USRA, NASA Ames Research Center, MS 232-12, Building N232, Rm. 146, P. O. Box 1, Moffett Field, CA 94035-0001, U. S. A.}
\altaffiltext{2}{Google, 1600 Amphitheatre Parkway, Mountain View, CA 94043, U.S.A.}
\altaffiltext{3}{Astronomy Department, University of California, Berkeley, CA 94720-3411, USA}
\altaffiltext{4}{UJF-Grenoble 1 / CNRS-INSU, Institut de Plan{\'e}tologie et dÕAstrophysique (IPAG) UMR 5274, Grenoble, 38041, France}
\altaffiltext{5}{Leiden Observatory, Leiden University, PO Box 9513, 2300 RA, Leiden, The Netherlands}
\altaffiltext{6}{Institute for Astronomy (IfA), University of Hawaii, 2680 Woodlawn Dr., Honolulu, HI 96822, USA}
\altaffiltext{7}{Centro de Astrobiolog{\'i}a$-$Depto$.$ Astrof{\'i}sica (CSIC/INTA), ESAC Campus, P.O. Box 78, 28691 Villanueva de la Ca{\~n}ada, Spain}
\altaffiltext{8}{Calar Alto Observatory, Centro Astron\'omico Hispano Alem\'an, C/Jes\'us Durb\'an Rem\'on, E-04004 Almer\'{\i}a, Spain}
\altaffiltext{9}{ALMA SCO, Alonso de C{\'o}rdova 3107, Vitacura, Santiago, Chile}
\altaffiltext{10}{Dep. de F\'isica Te\'orica, Fac. de Ciencias, UAM  Campus Cantoblanco, 28049 Madrid, Spain}
\altaffiltext{11}{Exoplanets \& Stellar Astrophysics Laboratory, NASA Goddard Space Flight Center, Code 667, Greenbelt, MD, 20771, USA}
\altaffiltext{12}{Eureka Scientific, 2452 Delmer, Suite 100, Oakland CA 96002, USA}
\altaffiltext{13}{Goddard Center for Astrobiology, NASA Goddard Space Flight Center, Greenbelt, MD 20771, USA}
\altaffiltext{14}{Kapteyn Astronomical Institute, Postbus 800, 9700 AV Groningen, The Netherlands}
\altaffiltext{15}{UMI-FCA (UMI 3386: CNRS France, and  U de Chile / PUC / U Conception), Santiago, Chile}

\begin{abstract}
The {\it Herschel Space Observatory} was used to observe $\sim$ 120
pre-main-sequence stars in Taurus as part of the
GASPS  Open Time Key project.  PACS  was used to  measure
the continuum as well as  several gas tracers such as [\ion{O}{1}]  63 
$\mu$m,  [\ion{O}{1}] 145 $\mu$m, [\ion{C}{2}] 158 $\mu$m, OH, H$_2$O
and CO. The strongest line seen is  [\ion{O}{1}] at 63 $\mu$m. We 
find  a clear correlation between the strength of the [\ion{O}{1}] 63 $\mu$m
line and the 63 $\mu$m  continuum for disk sources. In 
outflow sources, the line emission can be up to 20 times stronger than
in disk sources, suggesting that  the line emission is
dominated by the outflow. The tight correlation  seen for disk sources
suggests that the emission arises from the inner disk ($<$ 50 AU) and
lower surface layers of the disk where the gas and dust are coupled.
The [\ion{O}{1}]  63  $\mu$m is fainter in transitional stars than in normal Class II disks. Simple SED
models indicate that the dust responsible for the continuum emission is colder in
these disks, leading to weaker line emission. [\ion{C}{2}] 158 $\mu$m emission is
only detected in strong outflow sources. The observed line ratios of
[\ion{O}{1}] 63 $\mu$m to [\ion{O}{1}] 145 $\mu$m are in the regime where we are
insensitive to the gas-to-dust ratio,
neither can we discriminate between shock or PDR emission. We detect no
Class III object in [\ion{O}{1}]  63  $\mu$m and only three in continuum, at least one of which is a candidate debris disk.
\end{abstract}


\section{Introduction} 

Low-mass stars are born with protoplanetary disks, canonically composed
of 1$\%$ dust by mass (adopted from the ISM gas/dust mass ratio), the
remaining portion in gas.  For low-mass stars the gas has largely disappeared on timescales of 6 -- 7 Myr 
\citep{Haisch01,Hernandez08}.
Understanding the early evolution of stars requires the understanding of
their accompanying protoplanetary disks of gas and dust.

Gas in Protoplanetary Systems (GASPS) is a large Open Time Key Project
on the {\it Herschel Space Observatory} \citep{Pilbratt10} studying the
evolution of gas in protoplanetary disks. In total the project surveyed
$\sim250$ nearby ($\le$ 200 pc) low and intermediate mass stars (0.3 to
8 $M_{\sun}$), from young (0.5 Myr) stars with massive disks to older
(30 Myr) stars with very little dust excess. For this study we are using
the Photodetector Array Camera and Spectrometer (PACS, Poglitsch et al.
2010) to observe the fine-structure lines of [\ion{C}{2}] at 158 $\mu$m
and [\ion{O}{1}] at 63.2 and 145 $\mu$m as well as several H$_2$O, OH,
and high rotational transitions of CO  tracing mainly hot gas. We also
use the PACS imager to measure broadband continuum fluxes for a large
portion of the sample.  The goal is to understand the transition from
gas rich protoplanetary disks to gas poor debris disks, and the
timescales on which the gas dissipates.  For a more detailed description
of GASPS, see \citet{Dent12}.

The largest sample of GASPS sources are in the Taurus star forming
region.  As one of the closest star forming regions to the Sun (distance
of $\sim140$ pc), the Taurus group provides a unique opportunity to
obtain a statistical sample of largely coeval protoplanetary disks
encompassing a wide range of disk masses.  The majority of GASPS sources
in Taurus are class II objects (Classical T Tauri Stars, CTTSs, with
spectral types ranging from mid-F to early M, and with stellar masses of
$\sim$ 0.3 to 1.4 $M_{\sun}$), having dissipated their envelope of
accreting material and leaving behind the protostar and a disk of gas
and dust \citep[see e.g.,][]{Kenyon08}.  Of all the gas probes used by
GASPS, the [\ion{O}{1}] 63 $\mu$m line is by far the most sensitive gas
tracer, at least for protoplanetary disks \citep{Pinte10}.

The [\ion{O}{1}] 63 $\mu$m line is known to be strong in
photodissociation regions \citep{Tielens85}, and also in shocks
\citep{Hollenbach89,Flower10}. Because of this, it is necessary to
isolate the component of the emission arising from the disk itself from
that arising from jets or outflows, which are common in young T Tauri
stars. The velocity resolution of PACS is insufficient to separate the
low velocity disk emission from the outflow emission and for most
objects the outflow is spatially unresolved in  [\ion{O}{1}] 63 $\mu$m.
\citet{Podio12} have spatially resolved the [\ion{O}{1}] 63 $\mu$m
emission in a few of the strongest jet sources in the Taurus sample. In
our analysis we therefore distinguish between sources that are already
known to have jets/outflows based on previous observations and sources
which have very weak or undetectable outflows; in the latter the
[\ion{O}{1}] emission, if detected, is expected to be dominated by
emission from the protoplanetary disk. The relative importance of FUV
and X-ray irradiation for the [\ion{O}{1}] line emission from the Taurus
disk sources is explored in a forthcoming paper by Aresu et al. (2013,
in preparation). Detailed SED modeling and molecular emission lines will
be discussed in a future paper.

In Section~\ref{sec:Taurus} we discuss the Taurus sample, In
Section~\ref{sec:Obs} we discuss the observations and reduction for both
spectroscopy and photometry as well as pointing issues that arose in
some observations.  Section~\ref{sec:ResultsSpec} discusses the results
for the [\ion{O}{1}] and [\ion{C}{2}] fine structure lines, followed
with a discussion of what we learned in Section~\ref{sec:discussion}.
Section~\ref{sec:ClassIII} gives a brief summary of our results for
Class III objects and in Section~\ref{sec:individual} we comment on a
few  individual objects.  Section~\ref{sec:summary} gives the summary
and conclusions of this study.

\section{The Taurus-Auriga dark young stellar association}
\label{sec:Taurus}

The Taurus-Auriga cloud complex is one of the closest \citep[140
pc,][]{Bertout99,Torres09} active star forming regions, with a young
stellar association containing at least 250 cluster members
\citep{Rebull10,Rebull11}. It  does not have any O and B stars, but it has a rich
population of pre-main sequence low mass T Tauri stars and brown dwarfs.
The Taurus-Auriga dark cloud complexes extend over $\sim$ 10\degr\  of
the sky with a depth of $\sim$ 20 pc \citep{Torres09}. The youngest
stars, Class I protostars, are found toward the center of opaque dark
clouds. Likewise the CTTSs (Class II) are also concentrated towards dark
clouds, whereas the Class III objects, weak-lined T Tauri stars, are
more widely distributed over the whole region. Our Taurus sample also
includes 9 confirmed or suspected transitional disk objects.
Transitional disks, which were first discovered by \citet{Strom89}, are
believed to be  transition objects  between CTTSs and weak-lined T Tauri
stars. They have large mid- to far infrared excesses, but weak
(pre-transitional disks) or no excess in the near infrared
\citep{Najita07,Espaillat11}. Both pre-transitional and transitional
disks have large gaps or cavities largely void of gas and dust in the
inner disk. which in several cases have been resolved with millimeter
interferometry \citep{Pietu06,Dutrey08,Brown09,Andrews11}. The majority of the pre-main sequence
population in Taurus has an age of 1 - 3 Myr,
although there is certainly an age spread. Some stars are as young as
0.5 Myr and there are a few objects as old as 15 -20  Myr
\citep{Hartmann03,Kucuk10}.

Some of the younger, more deeply embedded T Tauri stars drive
Herbig-Haro (HH) flows or  well-collimated  jets, typically seen in the
optical [\ion{O}{1}], H$\alpha$, [\ion{N}{2}] $\lambda\lambda$ 6548,
6583 \AA, and [\ion{S}{2}] $\lambda\lambda$ 6717, 6731\AA\ emission
lines \citep{Mundt83,Gomez97}.  Since we try to discriminate between
disk emission and outflow emission in our GASPS observations, it is
therefore important to identify which stars power optical jets or HH
flows. For this we have mainly used the tabulation provided by
\citet{Kenyon08}, but we have also identified a few additional ones: 
AA Tau, DL Tau, and V773 Tau, see Section~\ref{sec:individual}, 
IRAS~04385$+$2805 \citep{Glauser08}, and SU~Aur \citep{Chakraborty04}. 
Stars that we define as jet/outflow
sources must have a jet  imaged in H$\alpha$,  [\ion{O}{1}]
$\lambda$6300 \AA,  [\ion{S}{2}] $\lambda$6371 \AA\ or be associated
with HH objects, show a high velocity molecular outflow, or a broad ($>
$ 50 km~s$^{-1}$), typically blue-shifted, emission line profiles in
[\ion{O}{1}] $\lambda$6300 \AA\ \citep[see e.g.][]{Hartigan95}.

For the most part, our analysis leaves aside the issue of multiplicity.
About half of our targets are known binary or higher-order multiple
systems, with separation ranging from $\lesssim 0.1$\,AU to as much as
2000\,AU. On an individual object basis, the interpretation of our
observations can be severely affected by the possible presence of
several circumstellar and/or circumbinary disks. However, as discussed
in Section~\ref{sec:mult}, it appears that the effect of multiplicity on
both far-infrared continuum and line emission is relatively modest.

\section{Observations and Data Reduction}
\label{sec:Obs}

The observations reported herein were carried out with {\it Herschel}
from February 2010 through March 2011.  We obtained photometry and
spectroscopy utilizing the PACS instrument aboard {\it Herschel}.  For a
description of PACS observing modes, see \citet{Poglitsch10}.  

\subsection{Photometry} 
\label{sec:phot}

All photometric observations of our targets in Taurus were made in
``mini-scan'' mode, i.e., by doing short,  3\arcmin\ scans with medium
scan speed of 20$^{\prime\prime}$/s, and with ten 4$^{\prime\prime}$
steps in the direction orthogonal to the scan. Executing such an
observation takes 276 seconds. Two bands, either blue and red, or green
and red, are recorded simultaneously.  The blue filter is centered on 70
$\mu$m, the green filter is centered at 100 $\mu$m and the red at 160
$\mu$m (see e.g. Poglitsch et al. 2010).  All targets have been observed
in the 70 $\mu$m and the 160 $\mu$m bands and many were also observed in
the 100 $\mu$m filter (see Table \ref{tbl-cont}).  Almost all the Taurus
photometric observations were carried out in two scan directions
(70$^{\circ}$ and 110$^{\circ}$) providing a smoother background and
hence more accurate photometry. However, some of the early photometry
was done with only a single blue/red mini-scan. Only T Tauri is
spatially resolved at 70 $\mu$m\footnote{The
emission from UZ Tau is also extended, but here the emission originates
from two binaries UZ Tau E and UZ Tau W, which are separated by 3\farcs6
and which both are surrounded by disks, see e.g. \citet{Harris12}}.
An extended disk is,  however,  detected around HD~283759, an F5 star, but
proper motion studies \citep{Massarotti05} show conclusively that the
star is not a member of the Taurus association.

All the photometric data were originally reduced using the {\it
Herschel} Interactive Processing Environment, HIPE\footnote{HIPE is a
joint development by the Herschel Science Ground Segment Consortium,
consisting of ESA, the NASA Herschel Science Center, and the HIFI, PACS
and SPIRE consortia.} v7.0, but we later re-reduced all detected sources
in HIPEv9.0, where the reduction scripts provide iterative masked high
pass filtering, resulting in better deglitching and smoother
backgrounds, especially in the red. For the blue filter the results were
essentially identical, i.e.,  within 1\% -- 2\%. The final images were
imported to the STARLINK program Gaia for photometry and further
analysis. For strong sources we performed aperture photometry with a
12$^{\prime\prime}$ aperture, while we used  6\arcsec\ - 9\arcsec\
apertures for faint sources and applied the aperture corrections 
determined by the PACS team \citep{Muller11}.  An annulus extending from
60$^{\prime\prime}$ - 70$^{\prime\prime}$ was used for estimating the
sky background. In some cases, where stars were near the edge of the
image, where we had multiple stars, or where there was extended
background emission in the image, the sky emission was estimated from
the average of several, ``clean'' areas in  the image. For the 70 $\mu$m
and 100 $\mu$m bands the background is very close to zero.  In a few
cases the star was surrounded by emission from the surrounding cloud.
This emission is mostly seen at 160 $\mu$m, but can be visible in the
100 $\mu$m band. In these cases we used a 6\arcsec\ - 8$^{\prime\prime}$
aperture, with the appropriate aperture correction applied. The point
source calibration for the PACS images has an absolute accuracy of 3\%
in the 70 $\mu$m and 100 $\mu$m band and better than 5\% at 160 $\mu$m 
\citep{Muller11}.  In our analysis we do not reach such accuracy. Many
of our target stars are embedded in or are in the direction of dark
clouds, and variations in the background limits the photometric
accuracy. We therefore estimate our photometric accuracy to be about
10\%.  In the 70 $\mu$m and 100 $\mu$m bands the 3-$\sigma$ upper limits
are typically about 9 mJy. In the 160 $\mu$m band the noise is much
higher due to the presence of ``cirrus''-like emission from dust clouds.
 Typical 3-$\sigma$ upper limits are about 20 mJy. All flux densities
are given in Table~\ref{tbl-cont} as measured, i.e., without color
corrections. The color corrections, however, are small ($<$ 2 \%) for
all our targets and do not affect our analysis. We also list the SED
classes in Table~\ref{tbl-cont} and Table~\ref{tbl-lineflux}. The SED
classification comes primarily from \citet{Luhman10} and or
\citet{Rebull10}; if they differ we list both. Some weak line T Tauri
star were not observed by \citeauthor{Luhman10} or
\citeauthor{Rebull10}. If we found no indication for an infrared excess
we assigned them as Class III. Transitional disks are labeled T.

Some fields also show faint unrelated sources. Almost all of them are
extragalactic background sources. A few, however, do coincide with faint
2MASS and WISE sources, and could be low-mass T Tauri stars. We 
therefore also list the number of ``background'' sources we find in each
field in Table~\ref{tbl-cont}. 

\subsection{Spectroscopy}
\subsubsection{Observations}

We observed the Taurus sample with PACS using both chopped line and
range spectroscopy modes, targeting primarily  [\ion{O}{1}] 63 and 145 $\mu$m and the
[\ion{C}{2}] 158 $\mu$m lines, but also the CO, OH, H$_2$O, CH$^+$, and  DCO$^+$
lines. Table \ref{tbl-spec} shows the setup for both line spectroscopy (henceforth
linespec) and range spectroscopy (rangespec). The primary lines are highlighted in bold.
The effective spectral resolution is 0.020 $\mu$m at [\ion{O}{1}] 63 $\mu$m  (88
km~s$^{-1}$) and 0.126 $\mu$m for the [\ion{C}{2}] 158 $\mu$m line (239 km~s$^{-1}$).
Therefore the observed lines are unresolved in velocity even for the shortest wavelengths,
which have the best spectral resolution, except in a few strong outflow sources, 
where the lines can be broad \citep{Podio12}.

For this study, spectra of 76  targets in 91 linespec observations were
obtained, covering the  [\ion{O}{1}] 63 $\mu$m and DCO$^+$ 190 $\mu$m lines (Table
\ref{tbl-obs}).  The integration times for the majority of observations was 1252 seconds; a
few targets were re-observed with longer integrations ranging from 3316 s to 6628 s.  Of
the 76 targets observed in linespec mode, 38 were also observed in rangespec mode (42
observations), with integrations of 5141--10279 seconds, covering 70 - 200 $\mu$m,
including the  [\ion{O}{1}] 145 $\mu$m and  [\ion{C}{2}] 158 $\mu$m lines. The
spectroscopic data were reduced with HIPE
\citep[HIPE,][]{Ott10} v4.2.0, utilizing the pipeline scripts provided at the {\it
Herschel} data reduction workshop in January
2010\footnote{https://nhscsci.ipac.caltech.edu/sc/index.php/ Workshops/}.

For each observation, we extracted the mean of the two nod positions to obtain flux values
for each of the 25 spaxels per observation.  We binned the spectra at half the
instrumental spectral resolution (Nyquist sampling) for a given wavelength within HIPE,
and output standard ascii text files in order to extract continuum and line flux
measurements.  We utilized the standard IDL line fitting routine MPFITPEAK \citep{Markwardt09} to derive the
total line flux and the continuum flux at the observed wavelength of each line.  We fit a
line to the observed continuum, defined nominally as the region of $\sim$2-10 times the
instrumental FWHM from the rest wavelength for each line.  The error on the continuum is
defined as the standard deviation of the residual about the mean.   The reported line
fluxes are the integrated flux of a gaussian line fit, with errors calculated from the
errors on the values of line amplitude and width, as reported by MPFITPEAK.  For spectra
with no line detection we report 3$\sigma$ upper detection limits, where the 1$\sigma$
flux  is calculated by integrating a gaussian with height equal to the RMS of the
continuum and width equal to the instrumental FWHM.   Typical rms error values for continuum
levels in individual spaxel spectra are $\sim0.05$ Jy.  Typical line flux errors (rms) are
$\sim0.5-1\times10^{-17}$ W m$^{-2}$.

\subsubsection{Correcting PACS spectroscopy observations for pointing errors}
\label{sec:pointing}

The PACS instrument consists of an array of 25 `spaxels', arranged 5 x 5 in a roughly
square formation; see \citeauthor[for a detailed description of the instrument]{Poglitsch10}. Each
spaxel subtends $9.4^{\prime\prime}$x$9.4^{\prime\prime}$ on the sky.   Ideally,
observations are carried out with the celestial source being centered on the central
spaxel.  For a `well-pointed' observation, the largest flux measured in the spaxel array
will, by construction, be located in the central spaxel. In such a case, the total flux of
a point source can be estimated from the observed flux in the central spaxel by applying
an aperture correction (which is the ratio of the observed flux in the spaxel to the total
source flux, as estimated from the instrument PSF), as provided by the HIPE reduction
software.

\begin{figure}
\centering
\includegraphics[height=3.5in]{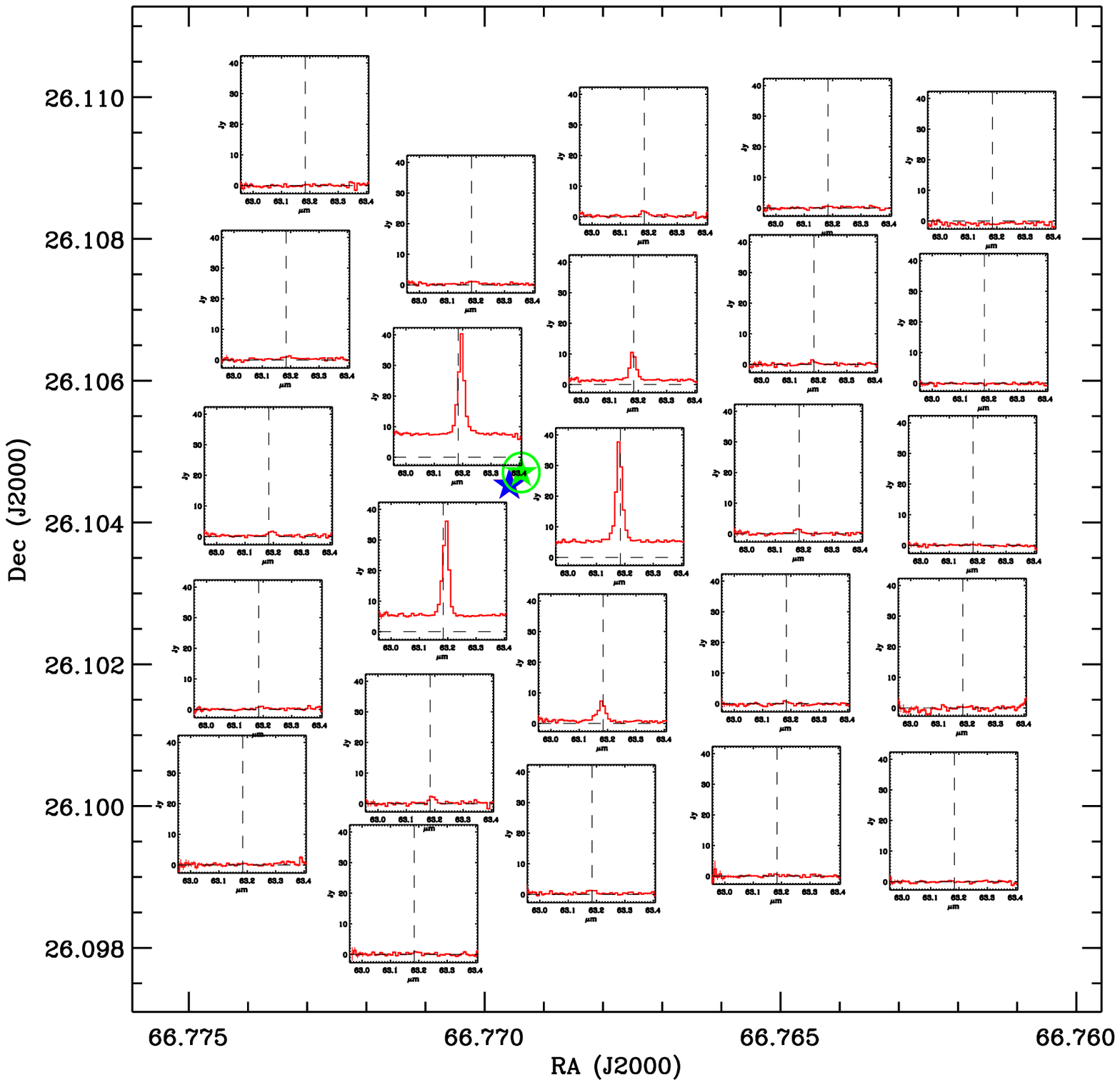}
\figcaption{``Footprint" for DG Tau.  Spaxels are plotted as $7^{\prime\prime}$ on a side
for clarity.  Because DG Tau is not extended in continuum at 63 $\mu$m, it is easily
identified as falling off the central spaxel by examining the relative continuum values in
each spaxel.  Plotted as the green star is the best fit position of the target via
$\chi^2$ minimization of spaxel continuum values and  the theoretical PSF at 63 $\mu$m.  A
$1^{\prime\prime}$ error circle representing the error in the positional fit is plotted. 
The blue star is the SIMBAD position for DG Tau.
\label{point}}
\end{figure}

However, some of the observations for our GASPS OT key program, especially for sources in
Taurus, were affected by large errors in the pointing. In some observations the sources were found to
be located up to $8^{\prime\prime}$ from the central spaxel, far greater than
reported pointing
accuracy\footnote{http://herschel.esac.esa.int/twiki/bin/view/Public/SummaryPointing} of
$\sim2^{\prime\prime}$.  As none of our Taurus sources are extended in continuum
emission (c.f. Section~\ref{sec:phot}), it is clear upon inspection of the relative
continuum fluxes in the spaxels of each observation if a source is not well centered
(Figure \ref{point}).  In the case of the GASPS'  Taurus targets,  the pointing solution
shows the  targets systematically to the East of the central spaxel.  For these
observations, the simple aperture correction method will not yield accurate results,
especially since the reduction software provides no way of determining precisely where the
source is located relative to the center of the spaxel array. Furthermore, the aperture
correction formula makes use of the signal (information) in a single spaxel and ignores
any observed flux in any additional spaxels. In order to determine accurately the flux of
our targets from these mispointed observations and overcome the shortcomings of the
aperture correction formula, we developed a method of estimating the offset of the source
from the center of the spaxel array, and using that information to calculate an ``optimum"
value of the source flux.

To do this, we simulated the observation of a source using wavelength-appropriate
theoretical PACS PSFs\footnote{http://pacs.ster.kuleuven.ac.be/pubtool.psf} and the known
spaxel sizes and locations relative to the central spaxel.  The theoretical PSFs are very
close to the observed PSFs \citep[see ][]{Poglitsch10},  deviating from the observed PSFs
on the order of only a few percent.  Relative spaxel positions on the sky  are obtained
from the RA and Dec coordinates provided for each spaxel in HIPE.  We take into
account the position angle of the telescope to transform the spaxel offsets from RA and
Dec to arcseconds in (y,z) spacecraft coordinates.

The simulations consisted of computing, for each spaxel,  the fractional flux as well as
the flux relative to the `observed' peak flux  as a function of offset in spacecraft
coordinates (y and z) between the peak of the PSF and the center of the spaxel array. We
stepped the spaxel array across the wavelength appropriate theoretical PSFs, generating a
library of spaxel fractional fluxes and relative fluxes for offsets of $\pm$ 20\arcsec, in
increments of 0\farcs5, in both directions.  For any given observation, the continuum fluxes
in each spaxel were divided by the peak continuum value and then compared to the simulated
relative fluxes at each offset position in the library using a $\chi^2$ statistic. The
positional offset (relative to the central spaxel) of the source is given by the offset
location that yielded the minimum $\chi^2$.  As in the line fits, the region used to
define the continuum in each spaxel was a range $\sim$2-10 times the instrumental FWHM
from the rest wavelength for each line.

Once the offset position was known, the fractional fluxes corresponding to that offset
were used to estimate the optimal value of the total source flux \citep[see e.g.][]{Horne86}. 
For the best fit offset position, we calculated an optimal flux value via an error
weighted average using only spaxels with a detected flux level above a signal to noise
ratio greater than 5.  The optimal estimate is given by \begin{equation}F_{opt.} =
\frac{\sum_{i=0}^n w_{i}\cdot f_{i}}{\sum_{i=0}^nw_{i}}\end{equation} with an error of
\begin{equation}\sigma_{opt.} = (\sum_{i=0}^n {w_{i}})^{-1/2}\end{equation}  The $w_{i}$
are weights given by \begin{equation}w_{i} = \frac{1}{(\sigma_{i}/p_{i})^2}\end{equation}
where $p_{i}$ is the fractional flux in spaxel $i$ as given by the previous determination
of the offset, and $\sigma_{i}$ is the measured statistical error (the standard deviation
on the mean continuum, as described above) on the continuum flux.   For bright sources,
several spaxels contribute to the final flux estimate, with each contribution weighted by
the fractional flux in that spaxel for the estimated offset position. For weak sources, or
those that are well-centered in a single spaxel, (and therefore, for which flux is
detected in only a single spaxel), the method should give the same result as the simple
aperture  correction formula.  A comparison of our `optimum' flux estimates and fluxes
using the single spaxel and the aperture correction value provided by HIPE, for
well-centered sources shows the values to be equivalent (i.e. for well centered sources,
or faint sources where flux is seen in only a single spaxel, the method of `optimal' flux
estimates described herein is consistent with standard flux estimates utilizing only the
HIPE reduction method).

For targets in which two sources were present in the PACS field of view (obsid 1342192193
containing FS Tau A/B, obsid 1342190351 containing XZ  and HL Tau,  obsid  1342192801
containing V807 Tau and  GH Tau), the method described above was modified somewhat to account
for additional flux from each source in each spaxel. The spaxels with the peak fluxes,
corresponding to the approximate locations of the sources, were identified. For each
individual source in the field of view, the method described above was carried out after
setting the weights for all spaxels between the two peaks and those associated with the
other source to zero. As above, the total flux of the sources was then determined from a
least-squares fit of the predicted fluxes in the remaining spaxels to the observed fluxes.
This yielded accurate relative locations, and thus accurate continuum flux values, of the
two sources in each field of view.

Testing of the code was performed on sources falling off the central spaxel (as
determined from "by-eye" inspection of relative continuum values) as  well as on well
centered sources.  The positional uncertainties in the y,z offsets, as defined by a
1$\sigma$ confidence contour based on the $\chi^{2}$ minimization, are  less than
$0.5^{\prime\prime}$ in radial offset, which is less than our step increments used in
generating the fractional fluxes falling in each spaxel.  As an additional check of the
offset code, we find very good agreement in derived offset positions between different
wavelengths (linespec and rangespec observations), which are separate observations, but
usually done back to back and which show the same pointing anomalies.  From a sample of 19
targets obtained in both spectroscopy modes (line scan and range scan), we find agreement
in derived radial offset positions computed using the 7 different wavelengths ranges, to
be better than $1^{\prime\prime}$ in radial offset.

As described above, all spectroscopy continuum values reported herein are the result of
the error weighted average of the aperture corrected flux in spaxels which have a S/N of 5
or greater.  Only one T Tauri star, T Tau itself, is extended  at PACS
wavelengths; Millimeter interferometry of many of the Taurus sources \citep{Andrews07,Isella09,Guilloteau11,Harris12} show
the disk diameters  to be $<$ 2\arcsec\ on the sky.  Since the observed sizes of the
objects in our sample will be smaller at the PACS wavelengths of 60--200 $\mu$m
than at millimeter wavelengths, any ``extended" continuum emission in multiple spaxels above and
beyond what would be expected from the PSF (i.e.,  adjacent spaxels with equal continuum
measurements) is a result of mis-pointing of the telescope.

However, the line emission can often be somewhat extended, especially for young  T Tauri
stars, as many of the sources in our sample are well studied outflow sources.  
In disks surrounding stars with known outflows, summing the line flux
contribution from all spaxels will overestimate the line emission arising from the disk,
since the spectral and spatial resolution of PACS at these wavelengths is insufficient to resolve the
outflow component of the line emission from that of the disk.  Therefore, the observed 
[\ion{O}{1}]63 $\mu$m emission in outflow sources only provides an upper limit to the emission
from  the disk. To minimize the contribution from the outflow, we only quote the line intensity from
the central spaxel (well-centered observations) or the expected line intensity at the continuum peak interpolated
from the line intensities in nearby spaxels. This is only significant for spatially extended sources like T Tau and  
DG Tau B, where our line intensities are about a factor of two smaller than the total (i.e., spatially integrated) line
intensities reported by \citet{Podio12}.

\subsubsection{Comparison of Spectroscopy and Photometry}
\label{sec:specPhot}

\begin{figure}
\centering
\includegraphics[height=2.2in]{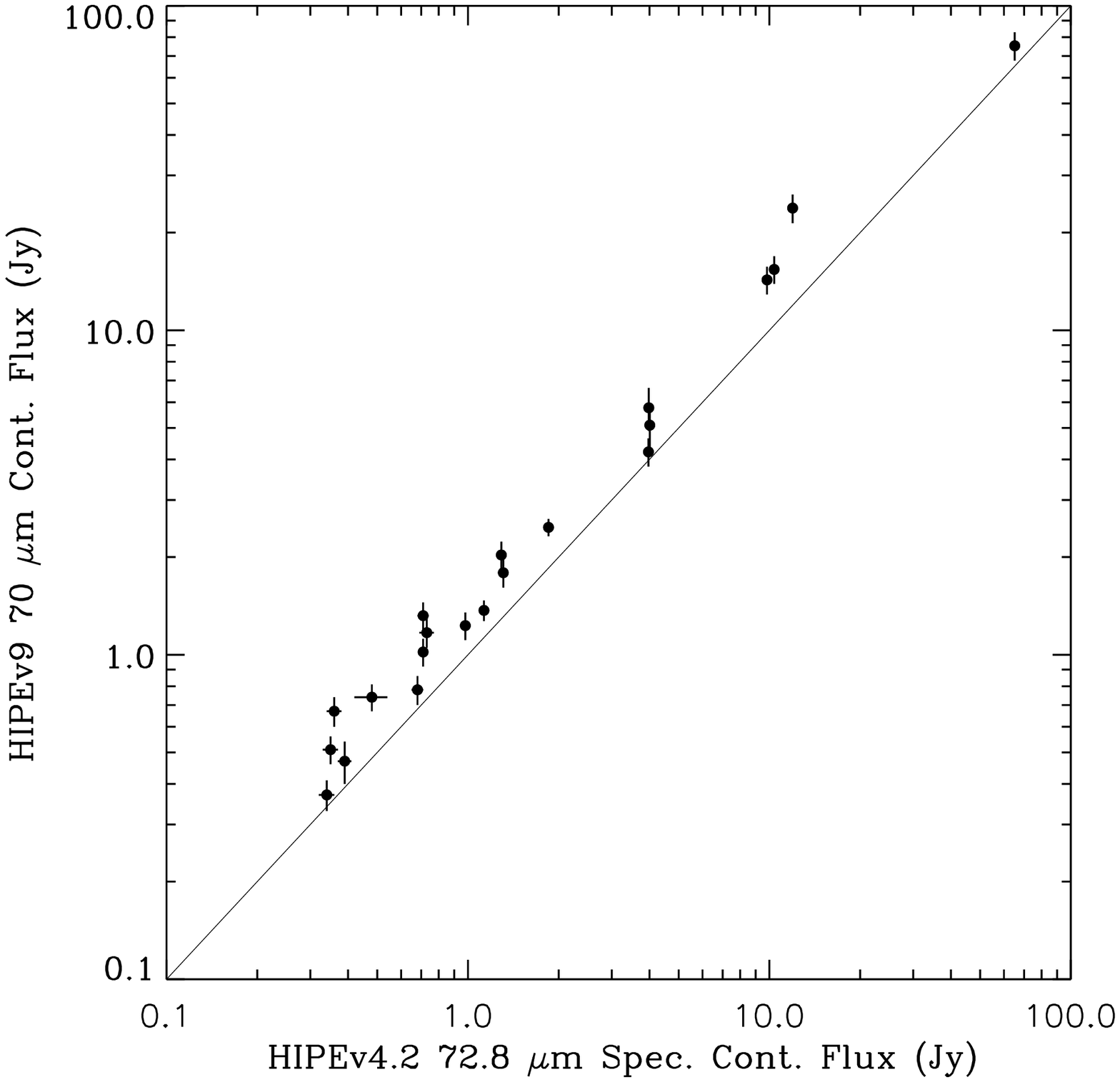}\\
\includegraphics[height=2.2in]{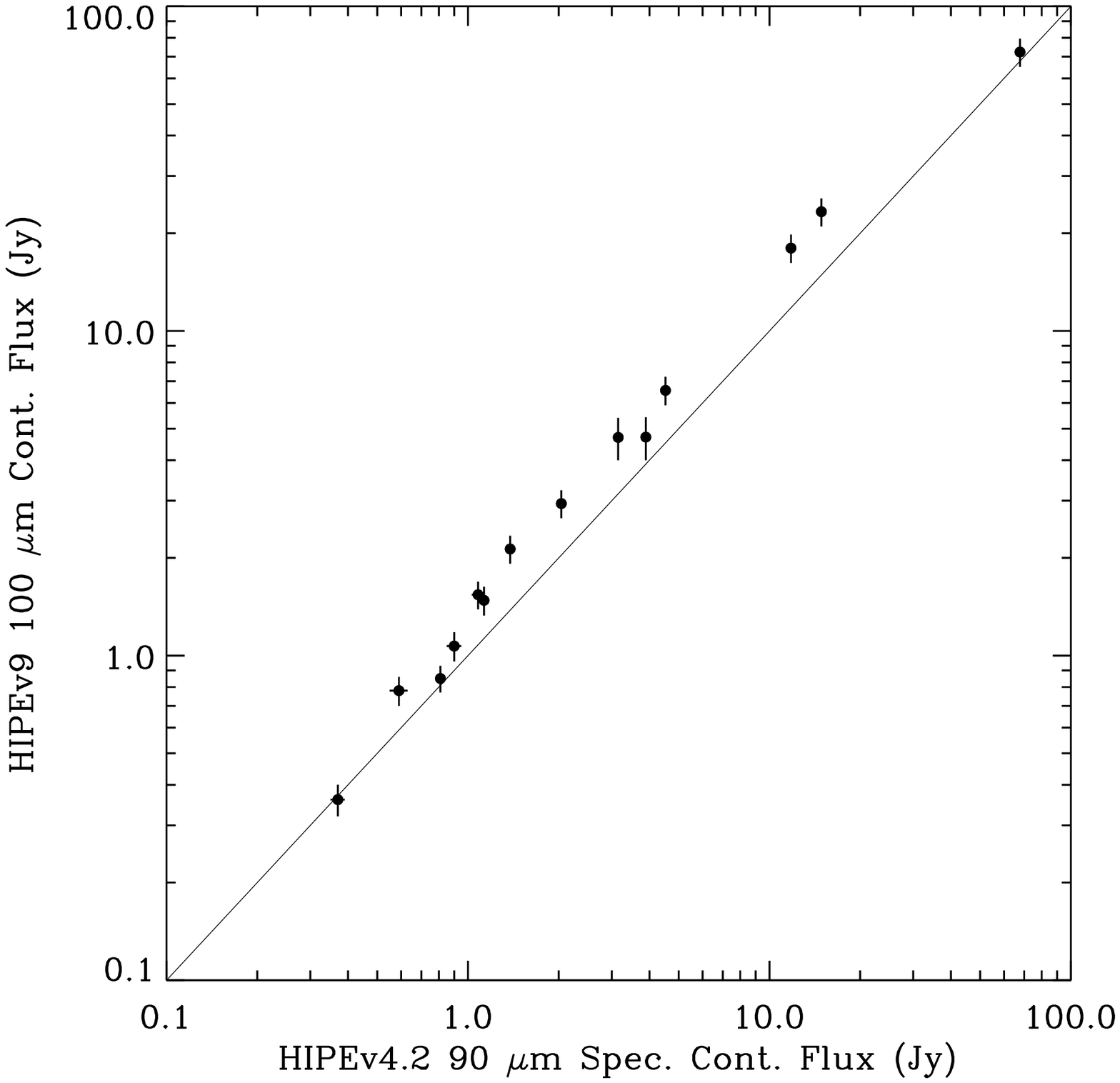}\\
\includegraphics[height=2.2in]{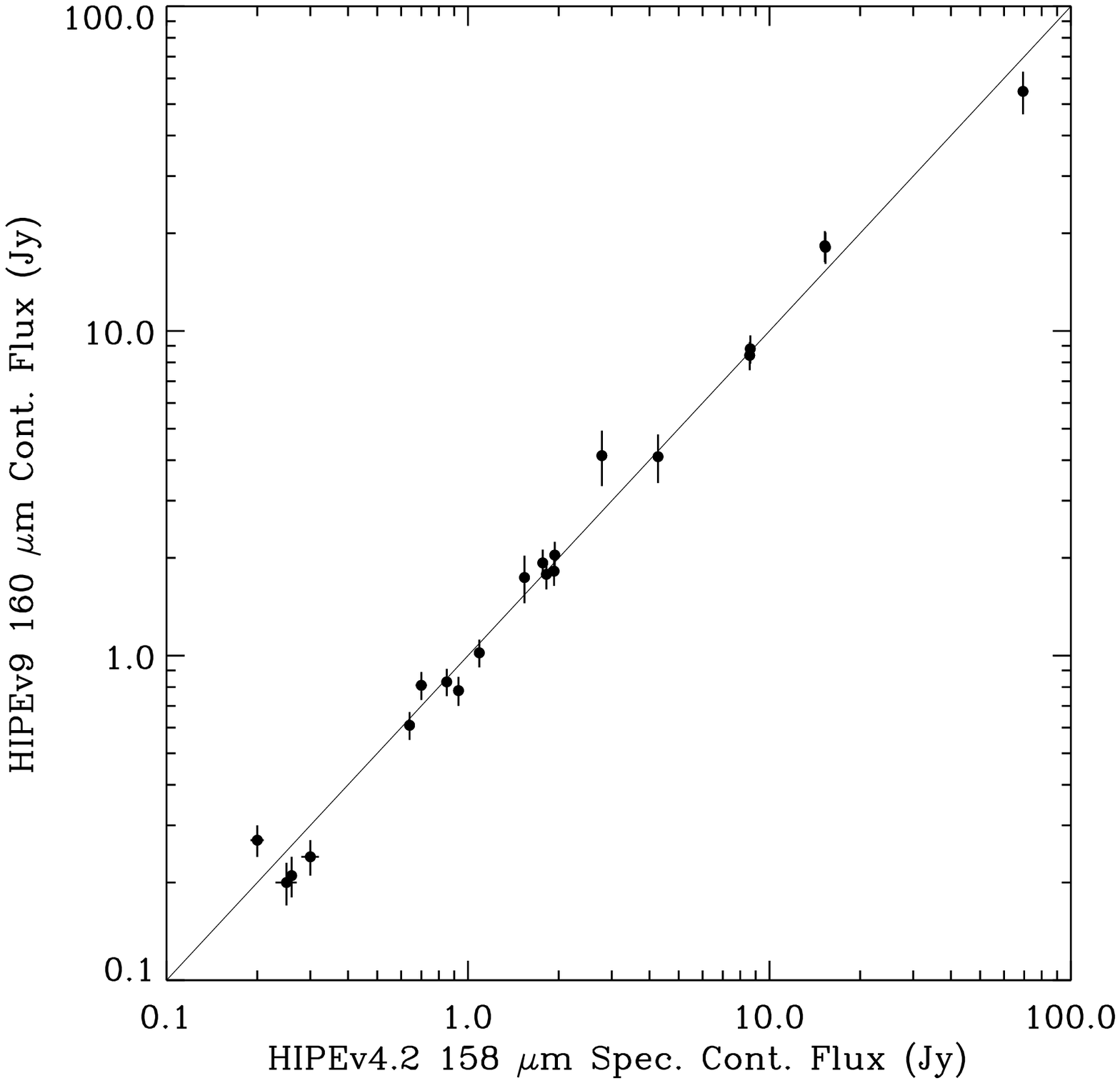} 
\figcaption{Comparison of PACS photometric values obtained with HIPEv9 and PACS
spectroscopic values obtained with HIPEv4.2.  Top:  70 $\mu$m photometry vs. 72.8 $\mu$m
spectroscopy. The photometric values at 70 $\mu$m are systematically $\sim$42\% (RMS
$\sim$26\%) higher than spectroscopy.  Middle:  At 100 $\mu$m the photometric values are
$\sim$33\% (RMS $\sim$20\%) higher than spectroscopy.  Bottom:  At 160 $\mu$m and the
photometric and spectroscopy values agree to within $\sim$2\% (RMS $\sim$20\%), on
average. 
\label{hipe}}
\end{figure}

We compare the continuum flux values obtained from photometry (Table~\ref{tbl-cont}) with
those from spectroscopy (Table~\ref{tbl-contflux}) in Figure \ref{hipe}. The photometric
calibration for PACS has been shown to have much higher absolute accuracy 
than the spectroscopic calibration \citep{Vandenbussche11,Muller11}.   We find that the photometry
continuum values are consistently higher by $\sim$ 30 - 40\% (RMS $\sim$20\%) than the spectroscopic values at both 70
and 100 $\mu$m.   This calibration discrepancy appears to be wavelength
dependent; comparison of the 160 $\mu$m photometric continuum and the 158 $\mu$m
spectroscopic continuum values show agreement to within 2\% (RMS $\sim$20\%), on average. 
We consider the absolute calibration of the spectroscopic data reduced with HIPE v4.2.0 to be accurate to within
$\sim$40$\%$.  

 In order to check how much the calibration has improved in the most recent HIPE release, v10.0, we
re-reduced the spectra at 63 $\mu$m for a randomly selected sample of 15 stars ranging from very faint, i.e. continuum
emission  $\leq$ 0.5 Jy at 63 $\mu$m, intermediate ( a few Jy),  to bright  (10 Jy) with about five stars in each category. The HIPE v10.0
calibration has certainly improved and is much closer to the results obtained from continuum photometry. From the comparison sample we find that the continuum flux densities are 18\% brighter in HIPE v10.0  than compared to HIPE v4.2,  while the line intensities are only 11\% brighter than in HIPE v4.2. Within errors, however, there is no clear difference between continuum and line calibration. For very faint sources HIPE v10.0 is no better than HIPE v4.2,  for both releases the spectroscopy calibration can be off by more than 50\% compared to
each other and to very accurate photometry\footnote{All spectroscopy and photometry will be re-reduced by the GASPS team as part of our data delivery once HIPE is stable. This is expected to change  the overall calibration for the spectroscopy bands B2A and B2B (50 $\mu$m -- 70 $\mu$m) by $\sim$ 20\%; for all other spectroscopy bands and for photometry the changes are expected to be small.}. 
Since we have overlapping targets with the DIGIT project \citep{Sturm10}, we also
compared the targets in common; the well-pointed DIGIT observations agree extremely well
with our photometry.   In general we find agreement to better than 15\% when comparing
spectra reduced in HIPE v4.2 to that of the DIGIT project, except for DG Tau, where the
discrepancy is 50\% (G. Herczeg 2011, private communication).

The spectroscopic data in this paper are presented as a self consistent dataset,
completely reduced with HIPE v4.2 in order to avoid confusion regarding scaling values for
targets reduced with different versions of HIPE.  As the main results of this paper focus
on comparison of continuum and line flux values at a given wavelength, the results should
be unaffected by an overall scaling of line and continuum flux values, except possibly for
line ratio values (discussed in Section \ref{sec:LineRatios}).  We note that in principle
both the continuum and line intensities should be scaled by the appropriate value as
determined by comparison to photometry for those wishing to use the continuum or line flux
values contained herein.  However, as there is not a 100\% overlap between sources observed
in photometry and spectroscopy modes, and because of the wavelength dependence of the
calibration, none of the spectroscopic continuum and line flux values (and
corresponding plots) in this paper has been scaled to match those obtained by  photometry.

\section{Results of Spectroscopy}
\label{sec:ResultsSpec}

Table~\ref{tbl-lineflux}  lists the  [\ion{O}{1}] 63 $\mu$m,
[\ion{O}{1}] 145 $\mu$m, and [\ion{C}{2}] 158 $\mu$m line flux values
for our GASPS Taurus sample. The PACS spectroscopy continuum values are
listed in Table \ref{tbl-contflux}. In some  of our targets, other 
molecular lines like OH, CH$^+$ and CO (for a complete listing, see
Table \ref{tbl-spec}) are also seen, primarily in known outflow sources
\citep[see, e.g.][]{Podio12}.  The o-H$_{2}$O line at 63.323 $\mu$m is
also detected in 8 of the sources in our sample
\citep{Riviere-Marichalar12}.  This water line is only seen in outflow
sources, although the emission is thought to originate in the disk
rather than in the outflow. Although discussion of these molecular lines
is outside the scope of this paper, targets which show evidence of these
various molecular lines are noted in Table \ref{tbl-lineflux}.  Here we
discuss only the strongest lines seen in the GASPS spectroscopic sample,
i.e. the two [\ion{O}{1}] lines and the  [\ion{C}{2}] line, which
dominate the gas cooling \citep{Tielens85}.

\subsection{[\ion{O}{1}] 63 $\mu$m emission}

The [\ion{O}{1}] 63 $\mu$m line was observed in 76 fields
(Table~\ref{tbl-obs}) and detected in {\bf 43} stars at 3$\sigma$ or more. It
was seen in all Class I objects, and in most Class II sources and
transitional disk systems. We found that we always detected the 
[\ion{O}{1}] 63 $\mu$m line at $\geq$3$\sigma$ if the 63 $\mu$ continuum
flux was $\gtrsim0.5$ Jy.  The line is seen in known outflow sources as
well as in ``disk-only" and transitional disk systems.  For outflow
sources,  the [\ion{O}{1}] 63 $\mu$m line emission is essentially
compact,  whereas the jets seen in H$\alpha$  or in forbidden optical
lines often trace the outflow over one or several arcminutes from the
star. The emission can be slightly extended in strong outflow sources
\citep[c.f.][]{Podio12}.  Line flux values in
our sample range from 10$^{-17}$ W m$^{-2}$ (our sensitivity limit) to
$\sim5\times10^{-17}$ W m$^{-2}$ for sources with no evidence of outflow
and up to almost 10$^{-14}$ W m$^{-2}$ for known outflow sources.  

\subsection{[\ion{O}{1}] 145 $\mu$m emission}

\begin{figure}
\centering
\includegraphics[width=5in]{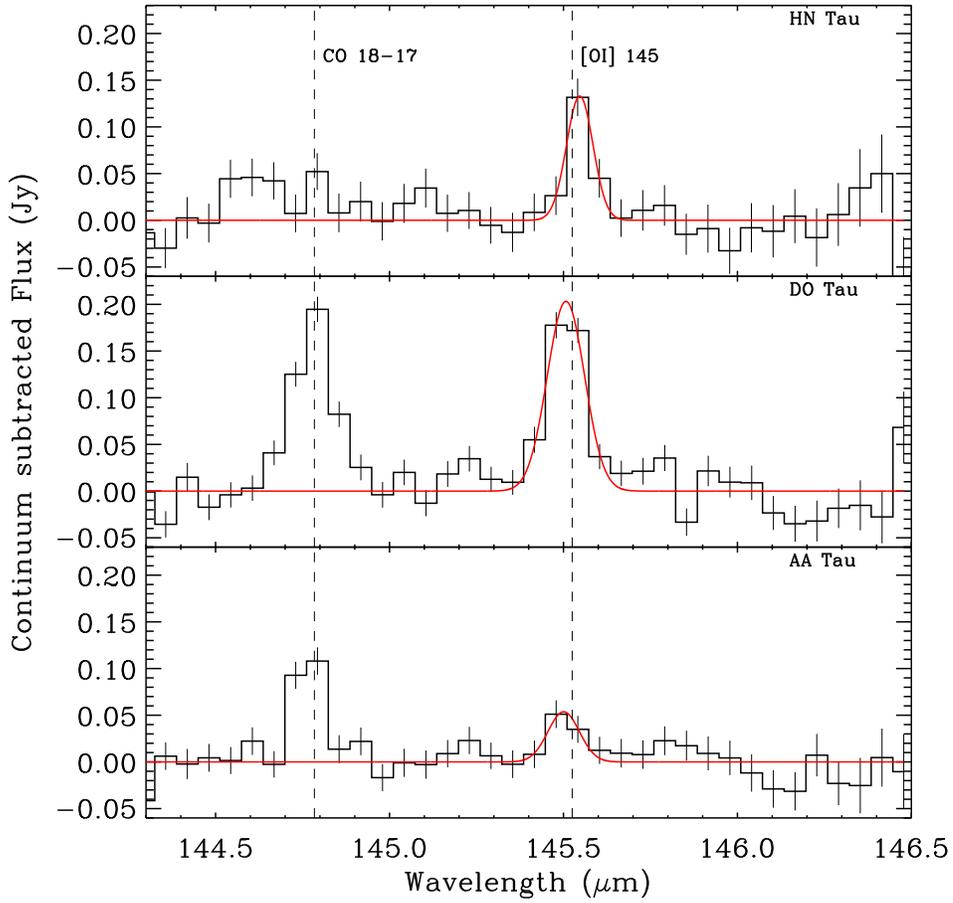}
\figcaption{ [\ion{O}{1}] 145  line detections for three of the targets in the Taurus
sample. The  line at 144.784 $\mu$m is  CO J = $18 \to 17$.
\label{OI145}}
\end{figure}

The [\ion{O}{1}] 145 $\mu$m line, with an energy level of $\sim$327 K, 
is more difficult to detect than the transition at 63 $\mu$m.  We detect
the 145 $\mu$m line at $\gtrsim$ 3$\sigma$ in 17 objects, nearly all of
them outflow sources (Table \ref{tbl-lineflux}); only one is a disk 
object. The non-detections of our disk sources show that the 
[\ion{O}{1}] 145 $\mu$m line is more than ten times fainter than the 
[\ion{O}{1}] 63 $\mu$m line in the disk. DE Tau is an exception, here
the 145 $\mu$m line is about as strong as the 63 $\mu$m line, see
Section~\ref{sec:individual}. This does not mean that the [\ion{O}{1}]
145 $\mu$m emission is absent in disks, but since our  disk sources are
relatively faint, the expected line emission is generally below our
detection limit, see Section~\ref{sec:LineRatios}. Figure \ref{OI145}
shows three examples of the [\ion{O}{1}] 145 $\mu$m line detections in
our sample.

\subsection{[\ion{C}{2}] 158 $\mu$m emission}

For the  [\ion{C}{2}] 158 $\mu$m line the detection rate is even lower than for
the  [\ion{O}{1}] 145  $\mu$m line; only 10 targets were detected, all of them outflow
sources.  Previous ISO
observations \citep{Creech-Eakman02,Nisini02,Liseau06} show strong ($\sim$10$^{-16}$--10$^{-15}$ W
m$^{-2}$) [\ion{C}{2}] 158 $\mu$m emission in young (mostly Class 0 \& Class I) stellar
objects. The [\ion{C}{2}]  is likely to originate in the outflow, rather than in the disk. In
the sample of strong outflow sources studied by \citet{Podio12}, they found that the 
[\ion{C}{2}] emission was somewhat extended compared to the continuum emission from the
disk for all but one source in their sample.  In the prototypical outflow source HH 46
observed with PACS,  \citet{van Kempen10} found that the  [\ion{C}{2}] emission was twice as
strong in the blue-shifted  outflow compared to the red-shifted outflow lobe or to the
spaxel centered on the protostar, demonstrating that  most, if not all of the  
[\ion{C}{2}]  emission originates in the outflow rather than in the disk.

For the outflow sources in our sample, we see [\ion{C}{2}] 158 $\mu$m line fluxes of
$\sim$10$^{-17}$ W m$^{-2}$ (DG Tau B and T Tau stand out with [\ion{C}{2}] 158 $\mu$m
line flux values of $\sim$10$^{-16}$ W m$^{-2}$), more than an order of magnitude lower
than the previous ISO observations. Because the ISO beam was very large at 158 $\mu$m,
$\sim$ 80$^{\prime\prime}$, it is likely that ISO picked up extended low surface
brightness emission from the surrounding cloud or emission from unrelated sources.

\subsection{Line Ratios}
\label{sec:LineRatios}

\begin{figure}
\centering
\includegraphics[width=4in]{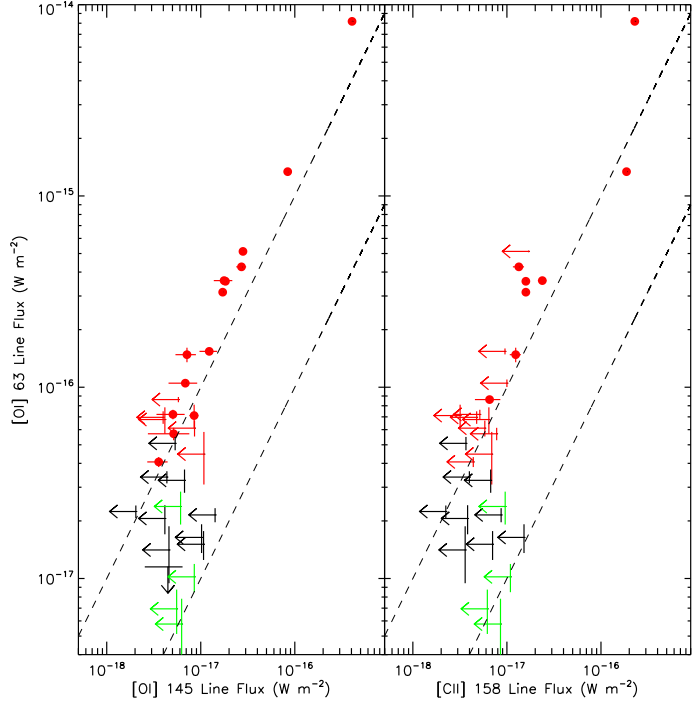}
\figcaption{ [\ion{O}{1}] 63 $\mu$m versus [\ion{O}{1}] 145 $\mu$m (left) and [\ion{C}{2}] 158 $\mu$m (right) 
line flux for the targets in the Taurus sample.  Solid circles are sources for which there
is a $\geq$3$\sigma$ detection for the lines.  Sources for which the line detection is
$<3\sigma$ are plotted as 3$\sigma$  upper limits. Red points are known outflow sources,
black non-outflow, and green are transitional disks.    The lower dashed line corresponds to a ratio
of one, and the upper dashed line to a ratio of 10.  Left:  All of the outflow sources show
ratios of $\sim$10 or higher, whereas the non outflow sources  are typically undetected in
[\ion{O}{1}] 145 $\mu$m, with the exception of DE Tau.
\label{ratios}}
\end{figure}

Figure \ref{ratios} shows the line ratios for  [\ion{O}{1}] 63 $\mu$m, [\ion{O}{1}] 145
$\mu$m, and [\ion{C}{2}] 158 $\mu$m.  In all cases where the lines are detected at
3$\sigma$ or higher, the line ratios are 10 - 25 for [\ion{O}{1}] 63/ [\ion{O}{1}] 145. The
line ratios for [\ion{O}{1}] 63/ [\ion{C}{2}] 158 are  similar. Due to the wavelength
dependence on calibration with spectroscopy (discussed in Section \ref{sec:specPhot}),
these ratios are likely to be even higher, as the [\ion{O}{1}] 63  $\mu$m line flux is
systematically underestimated in comparison to the [\ion{O}{1}] 145  $\mu$m and [\ion{C}{2}] 158  $\mu$m
line fluxes.

\subsection{63 $\mu$m Continuum versus [\ion{O}{1}] 63 $\mu$m emission}
\label{sec:cont_vs_63}

Since the  [\ion{O}{1}] 63 $\mu$m line can be strong in outflows, we
divided our sample into two sets: 1) the ``disk-only'' stars, where the
star has no or only weak outflow activity and 2) jet/outflow sources,
where there is clear evidence for outflow activity (see
Section~\ref{sec:Taurus}). In Figure~\ref{corr} we plot the [\ion{O}{1}]
63 $\mu$m line intensities versus the 63 $\mu$m continuum flux density. 
We find a tight correlation between 
[\ion{O}{1}] line emission and 63 $\mu$m continuum flux with a
correlation coefficient of 0.90 for the disk sources which have both line
and continuum fluxes $>$ 3$\sigma$. A weighted fit gives 
\begin{equation}log(OI) = (0.737  \pm 0.06) \times
log(S_{63}) - (0.22 \pm 0.02) \end{equation} where OI is the  [\ion{O}{1}] 63 $\mu$m line
flux in $10^{-16}$ W m$^{-2}$ and S$_{63}$ is the continuum flux in Jy
at 63 $\mu$m. This fit is shown as a black line in Figure~\ref{corr}.
Our typical 3$\sigma$ line detection limits, $\sim10^{-17}$ W m$^{-2}$,
are not stringent enough  to conclusively determine if the trend
continues to values below $\sim0.3$ Jy. All transitional disks lie below
the correlation. In order to verify this correlation is valid beyond the 12 detected disk-only sources, we have used Kendall's Rank Correlation, a correlation statistic that accounts for non-detections, but does not include uncertainties on detected sources \citep{Isobe86}.  We use an IDL adaptation of the algorithm presented in \citet{Isobe86} to carry out the calculation.  The population of 12 detected sources have a Z-value of 3.57, indicating there is a $3.6\times10^{-4}$ probability that a correlation is not present.  When all 40 disk-only objects are included, the Z-value increases to 5.29, indicating that the probability of no correlation is essentially 0.

Furthermore, we carry out a log-log fit to the fluxes and 3$\sigma$ upper limits of all disk-only sources.  This fit is a modified $\chi^2$ minimization using the Levenburg-Marquardt algorithm as implemented in the IDL code MPFIT \citep{Markwardt09}.  For detections, the contribution to $\chi^2$ is determined as normal (i.e. $((O - M) / \sigma)^2$).  For non-detections, the $\chi^2$ value is set such that if the model value is below the 4$\sigma$ limit, the point contributes a value of 1 to the deviates.  For values above 4$\sigma$, the $\chi^2$ value is calculated as if the objected were detected at the 3$\sigma$ value with an uncertainty of 1$\sigma$.  Parameter uncertainties from this procedure are lower-limits, and do not reflect the two-sided nature of the uncertainties.

This log-log fit results in a slightly shallower slope, 0.69 $\pm$ 0.04, but is still well within the 1$\sigma$ errors.  These tests of the correlation and the line-fit to the disk-only locus confirm that the line flux and the 63$\mu$m flux are strongly correlated.  

%
%

While the outflow sources show a similar trend, i.e., that sources with higher continuum flux tend to have higher line flux, they can exhibit line fluxes that are up to  $\sim$20 times higher than non-outflow sources for a given continuum
value. For these outflow sources, it is clear that the [\ion{O}{1}]  63
$\mu$m line emission is completely dominated by the outflow.  However,
there are some outflow sources,  which lie along the best-fit line to the disk-only sources (6
out of 24), suggesting that in some cases the shocks in the jets are not
strong enough to excite the [\ion{O}{1}]  63 $\mu$m. Nevertheless,  the
strong correlation between [\ion{O}{1}] emission and 63 $\mu$m continuum
for disk sources suggests that one may can use low spectral resolution
[\ion{O}{1}] observations  to determine whether a PMS star drives an
outflow or use the derived correlation to provide a rough estimate of
how much of the  [\ion{O}{1}]  63 $\mu$m line emission originates in the
disk.

Several sources in our sample were  detected in continuum emission at 63
$\mu$m but showed little (below 3$\sigma$) or no [\ion{O}{1}] 63 $\mu$m
line emission in our first set of observations.  Some of the sources 
for which there was a hint of a line were re-observed with roughly three
times the exposure, yielding line detections on the order of
$0.5-1\times10^{-17}$ W m$^{-2}$ as shown in Figure \ref{corr}. All the sources, which were re-observed, were detected.  For
several re-observed sources, the line flux is lower than what the
correlation would predict. Some of these sources, however,  are known
transitional systems (DM Tau, DN Tau, UX Tau A, GM Aur,  and LkCa 15),
which we already identified as having weaker  [\ion{O}{1}]  line
intensities than CTTSs for the same continuum flux. This result suggests
that the correlation between line flux and continuum is not only a good
indicator of whether a source drives an outflow, but can also serve as a
diagnostic for transitional disks. These transitional disks will be
discussed in detail in a forthcoming paper (M\'enard et al. 2013, in 
preparation).

If we include marginal line detections, i.e.,  $>$2$\sigma$ but
$<$3$\sigma$, we find an additional 7 sources which lie below the
correlation for ``disk-only'' sources. Four  of these marginal
detections are transitional disks \citep[FO Tau, CX Tau, IP Tau, and
V836 Tau;][]{Najita07,Espaillat11,Furlan11},  which confirm the trend we
already saw, i.e. transitional disks are intrinsically fainter in
[\ion{O}{1}] 63 $\mu$m than classical disks. The remaining three (Haro
6-37, FM Tau, and V710 Tau; see Table~\ref{tbl-lineflux}) may be entering
the transitional phase of their evolution, or alternatively they appear
faint due to increased systematic errors for faint sources. 

One other source stand out, GH Tau. It has a continuum flux density of 0.8 Jy at
63 $\mu$m (Table~\ref{tbl-contflux}), yet the  [\ion{O}{1}] 63 $\mu$m
line emission  is less than 3$\sigma$. However, at 70 $\mu$m the flux
density is only 0.37 Jy (Table~\ref{tbl-cont}). Therefore the continuum
emission at 63 $\mu$m is overestimated by more than a factor of two, and
the  [\ion{O}{1}] 63 $\mu$m  2$\sigma$ detection is at the level we
would expect.

\section{Discussion}
\label{sec:discussion}

\subsection{[\ion{O}{1}] 63 $\mu$m emission}

The [\ion{O}{1}] 63 $\mu$m line, with an upper energy level of  228 K,
is  one of the strongest FIR lines observed in the PACS wavelength range
and a dominant cooling transition.  Emission is thought to arise from
the surface region of almost the entire disk \citep{Gorti11, Kamp11}
with 50\% of the emission coming from outside of 100 AU \citep{Kamp10}. 
\citet{Meijerink12} find the bulk of the atomic oxygen to be in LTE, and
hence the [\ion{O}{1}]  63$\mu$m  and 145 $\mu$m lines should be
sensitive to the average temperature of the emission regions they are
probing, making them potential probes of the energy of the gas. Several
models show that the [\ion{O}{1}] 63 $\mu$m  line is  optically thick
everywhere in the disk \citep{Gorti08,Kamp11}.

Thermo-chemical models of disks \citep{Gorti08, Kamp11} predict
much higher line fluxes than what we observe for our disk sources, even after accounting
for calibration uncertainties (our flux estimates are predicted to be $\sim$ 30\% low, see Section~\ref{sec:specPhot}),
suggesting that the [\ion{O}{1}] emission is still not fully understood.
Outflow sources  have much stronger line fluxes than their
disk-only counterparts, with values that can be up to 20 times stronger
for the same continuum flux compared to  a non-outflow source. 
Therefore an outflow, if present, can dominate the line
emission, especially for strong outflows.  This is discussed further in Section
\ref{sec:cont_vs_63}.

\subsection{What can we learn from the [\ion{O}{1} 63/145 Line Ratios?}

If the emission of [\ion{O}{1}] 145  $\mu$m and [\ion{O}{1}] 63  $\mu$m
came from the same region, our observed line rations 10 - 25,  would
suggest optically thin emission originating from gas with a temperature
of a few 100 K \citep{Tielens85} located at the inner part and surface
layers of the disk or from the shock regions in the outflow. However, we
would also see similar ratios for optically thick lines, if the 
[\ion{O}{1}] 145 $\mu$m line, which requires hotter gas, came from a
more compact source (like only the inner part of the disk).

Disk sources (DE Tau is an exception) and transitional disks do not show
 either [\ion{O}{1}] 145 $\mu$m or [\ion{C}{2}] 158  $\mu$m, because
these disks are intrinsically faint and the emission is below our
detection limit.

\citet{Kamp11} explored the  [\ion{O}{1}]  63/145 line ratios
using the results from the DENT grid (Disk Evolution with Neat
Theory) that consists of 300000 disk models with 11 free parameters.
They find that  the median of  the [\ion{O}{1}]  63/145 line ratios for the whole DENT
grid is 25, while the ratio is $\sim$ 16 for the canonical gas-to-dust
ratio of 100. Similar ratios are found for the disk
models by \citet{Meijerink08}, who found that [\ion{O}{1}] 145 $\mu$m
line emission is 20 -- 40 times fainter than the [\ion{O}{1}]  63 $\mu$m
line. This also agrees with the results by \citet{Meeus12}, who studied
a sample of isolated HAEBE stars. They found that when the  [\ion{O}{1}]
145 $\mu$m line was detected, it was typically 20--30 times fainter than
the [\ion{O}{1}] 63 $\mu$m line. For such line ratios we would not
expect to detect the  [\ion{O}{1}] 145 $\mu$m line in any of our
``disk-only'' sources. Line ratios of 10 -- 50 are also predicted by
shock models \citep{Hollenbach89,Flower10}. Thus the [\ion{O}{1}] 
63/145 line ratios does not help us to discriminate between a jet or a
disk origin.

\citet{Liseau06}, who analyzed all ISO-LWS observations of young objects
with outflows (mostly Class I and Class 0 objects), found surprisingly
low [\ion{O}{1}] 63 $\mu$m to [\ion{O}{1}] 145 $\mu$m ratios; more than
half  of their sample had line ratios less than 10.  The average line
ratio of the ones that had a line ratio less than 10 was  5.3 $\pm$ 2.6.
Even though our sample consists mostly of Class II sources, we do
include  some Class I sources and several of our Class II sources drive
powerful outflows. We would therefore have expected to see some low
ratios, especially since all the sources detected in [\ion{O}{1}] 145
$\mu$m are outflow sources, but we do not. Essentially all of our
detections have a ratio $\gtrsim10$, see Figure~\ref{ratios}. A recent
paper from the guaranteed time  Key program WISH \citep{Karska13} discuss  PACS observations of
[\ion{O}{1}]  in 18 low-mass protostars, 5 of which were Class I and the
rest were Class 0 sources. They also find much higher line ratios than
\citet{Liseau06}.  In their Class 0 sample \citeauthor{Karska13} found 
the [\ion{O}{1}] 63 $\mu$m to [\ion{O}{1}] 145 $\mu$m line ratios to
vary from 5.5 to $>$ 45, with a median of $\sim$ 10.4, while their Class
I samples varies from 10.4 to 26.5. Overall there does not seem to be
much of a difference between the results obtained by
\citeauthor{Karska13} and us. Their Class 0 sample has somewhat smaller
ratios but the ratios for the  Class I sources are about the same as we
find in Taurus. The most likely explanation  for the lower ratios found
by Liseau et al. is probably due to calibration. The difference in beam
size should have a rather minor effect, since the  [\ion{O}{1}] emission
is rather compact, and if anything one would expect the 63 $\mu$ line to
be spatially more extended than the 145 $\mu$m line, which would push
the ratio to higher values, not lower ones.

\subsection{Disk mass versus  [\ion{O}{1}] 63 $\mu$m line emission} 
\label{sec:dmass}

\citet{Kamp10} suggested that the  [\ion{O}{1}] 63 $\mu$m and the  [\ion{C}{2}]  158
$\mu$m lines, especially when combined with CO, may provide a tool to measure the disk gas mass \citep[see also,][]{Kamp11}.
 Our observations, however, show that  even with {\it Herschel} we do not have the
sensitivity to detect [\ion{C}{2}] in disks around low mass stars; all our  [\ion{C}{2}] 
detections are outflow sources and the line emission, when detected, is almost certainly
from shock or PDR emission in the outflow \citep{Podio12}. On the other hand,  [\ion{O}{1}] 63 $\mu$m, as 
shown in the previous section, clearly originates in the disk, because we detected the
line in  20 stars, which show little or no outflow activity. There are not enough observations of any given CO transition to enable a meaningful
investigation of how  the  [\ion{O}{1}]/CO ratio correlates with disk mass. However, if the  [\ion{O}{1}] 63 $\mu$m
emission probes the disk gas mass, one might expect  some sort of correlation between 
[\ion{O}{1}] line flux and total disk mass.

\begin{figure}
\centering
\includegraphics[width=6in]{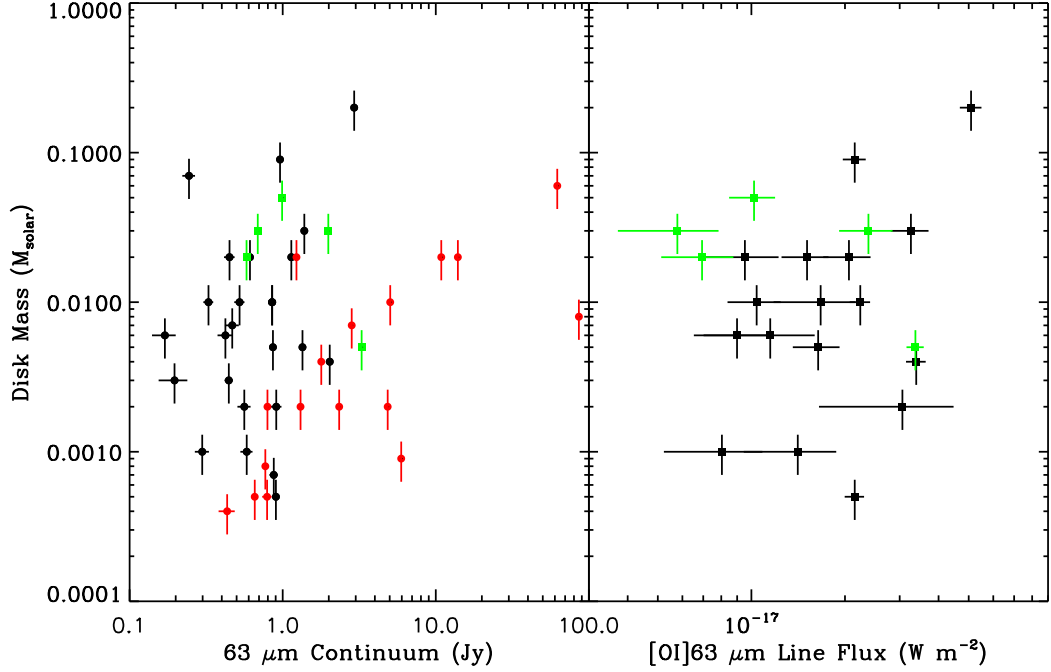}
\figcaption{Disk mass estimates from \citet{Andrews05} as a function of [\ion{O}{1}] 63 $\mu$m line flux and 63
$\mu$m spectroscopic continuum flux (right panel) and the 63 $\mu$m continuum (left panel).  Detections of $\lesssim3\sigma$ in either
continuum or line flux are omitted.  Red points are targets with known outflow activity,
black points are sources with no known outflow, and green points represent known
transitional disk sources.  Known outflow sources have been omitted from line flux plot, as
our analysis show that when a star drives an outflow, the outflow dominates the
[\ion{O}{1}] 63 $\mu$m  emission. There is no correlation between disk mass and [\ion{O}{1}] 63
$\mu$m line emission nor with the 63 $\mu$m continuum flux.
 \label{DMASS}}
\end{figure}

After searching through the literature, we decided to use
 the disk mass estimates for the large sample investigated by
\citet{Andrews05}. A homogeneous treatment is more important than accuracy 
for this investigation, because published mass estimates vary greatly
depending on how they have been estimated, and what dust emissivities 
have been used \citep[see e.g.,][]{Ricci10,Guilloteau11, Ricci12}. As we can see
from the right panel in Figure~\ref{DMASS}, there is no correlation
between  [\ion{O}{1}] 63 $\mu$m line flux and disk mass. We increased our sample by
plotting the disk mass as a function of 63 $\mu$m continuum flux
density, since the continuum emission originates in the disk and is not
affected by the outflow. However, as we can see from the left
panel in Figure~\ref{DMASS}, where we plotted outflow sources in red,
they also reveal no correlation and the same is true for the whole
sample. For the same 63 $\mu$m continuum flux density or line intensity
the disk mass can vary by more than a factor of a hundred.

\subsection{Source of 63 $\mu$m Continuum Emission}
\label{sec:ContSource}

Why is there a correlation between [\ion{O}{1}] 63 $\mu$m and the 63 $\mu$m continuum? The
simplest explanation for the tight correlation is that both line and continuum emission come from the same
region of the disk.  In order to investigate where the 63 $\mu$m continuum emission originates in
the disk, we first investigated the simple disk model used by  \citet{Andrews07}, which
uses the broadband SEDs and sub-millimeter visibilities to constrain some of the basic
parameters that describe the structure of the disk.  Using their flat disk model \citep[as
described in ][]{Andrews07}, we find that for the majority of their sources that are also
in our Taurus sample, 90\% of the 63 $\mu$m continuum emission comes from the inner
$\sim$10\%, or $\sim$5--50 AU, depending on the size of the disk.   It should be noted,
however, that overall disk size is poorly constrained in these models. Models of T Tauri stars with MFOST \citep{Pinte06}, which include disk flaring,
give similar results.  For the Class II objects in our Taurus
sample, it therefore appears  that a significant fraction of the line emission also originates in the inner  part of the disk.

\begin{figure}
\centering
\includegraphics[width=3.2in]{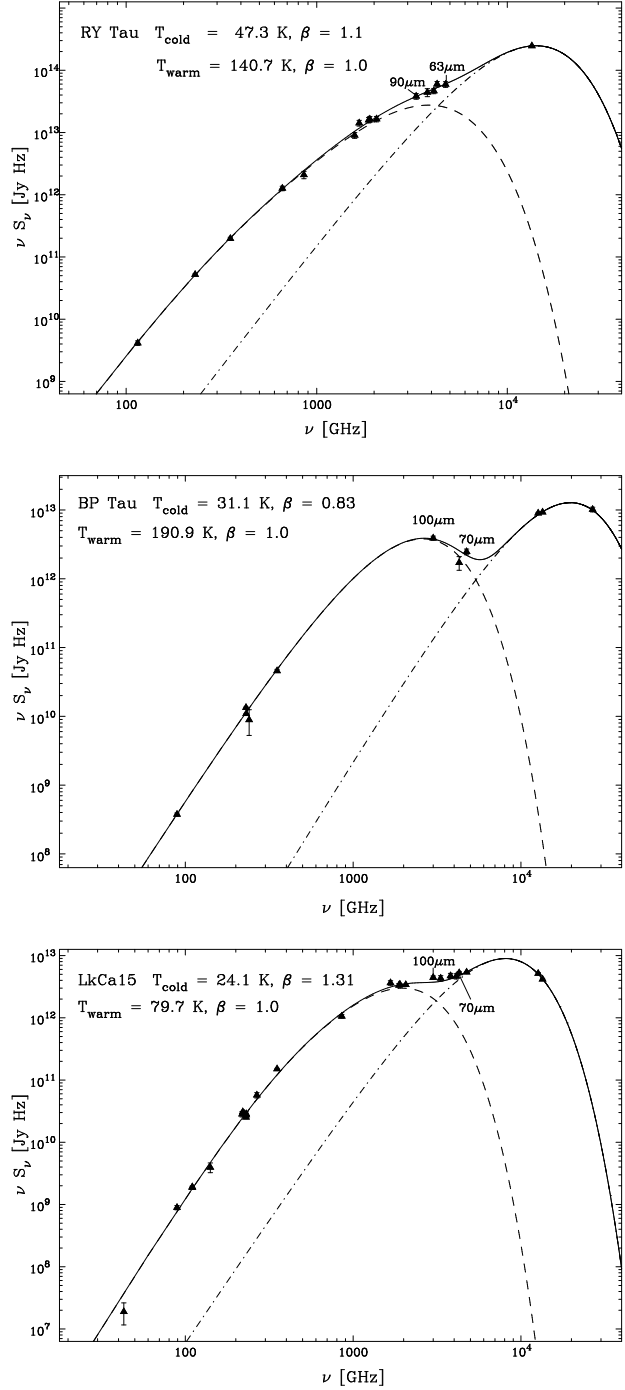}
\figcaption{Representative two component greybody fits. Millimeter and sub-millimeter data
come from literature, far infrared data from this paper, MIPS 24 and/or 70 $\mu$m fluxes
are also from literature, while WISE 22 $\mu$m fluxes are from the WISE All-SKy Catalog.
Since we do not  include photometry shortward of 12 $\mu$m, we can ignore the hot inner disk and the stellar photosphere.
\label{greybody}}
\end{figure}

As a next step we explored the bulk properties of the disks  using
isothermal  two-component grey body models. For these fits we use all
published millimeter and sub-millimeter continuum fluxes combined with
the far infrared fluxes from this paper. To constrain the warm dust in
the disk we also included MIPS 24 $\mu$m and/or WISE 22 $\mu$m data from
the WISE All-Sky survey. The MIPS 24 $\mu$m data are taken from
\citep{Luhman10} and/or \citet{Rebull10}; the latter also includes MIPS
70 $\mu$m data. The 22  $\mu$m and 24 $\mu$m mid-IR flux densities are
still completely dominated by dust emission from the disk for the Taurus
Class II  sources and the contribution from the photospheric emission is
negligible. The fitting program allows us to put in constraints for the
dust temperature, and fits separately the size of both the cold and the
warm dust as well as the dust emissivity index. In order to limit the
number of fitted parameters we do not fit the dust emissivity index,
$\beta$, of the warm component; instead we set it to 1. Whether we use
$\beta$ =1.0 or 1.5 does  not really matter, the dust temperature is
essentially the same to within a degree. The results of these fits for a
sample of 21 sources  (both disk and outflow sources, but not
transitional disks), for which we had sufficient data, indicate that the
temperature of the warm dust ranges from $\sim$ 95 K to 190 K, with a
median of 130 K. This is about the temperature we would expect for the
[\ion{O}{1}]  emitting gas (Uma Gorti, private communication). There is
no difference in the temperature of the warm dust between outflow
sources and non-outflow sources. Figure \ref{greybody} presents several
examples of these grey body fits.  These results suggest that  the 
[\ion{O}{1}]  emission originates in regions where the gas and dust are
approximately thermalized. This might be  the inner part of the disk or
the warm, lower surface layers of the disk.

As we saw earlier, we also see a correlation between  [\ion{O}{1}]  63
$\mu$m flux and continuum emission in the transitional disks, but the 
[\ion{O}{1}] emission is consistently weaker by a factor of two or more
for the same  continuum flux. Two component grey body fits of these
transitional disks yield  substantially cooler temperatures for the warm
dust, i.e. the dust dominating the 63 $\mu$m continuum emission. For
this sample the warm dust  covers a relatively narrow temperature range,
about 80 - 90 K, with a median  of 85 K.  This is consistent with the
lower observed  [\ion{O}{1}] 63 $\mu$m line emission in these
transitional disks. If the dust and gas is coupled in [\ion{O}{1}] 
emitting regions of the disk, the lower temperature compared to
``normal'' Class II disks readily explains the lower line intensity. We
note, however, that this interpretation may be oversimplified. More
detailed modeling of transition disks will be presented in a forthcoming
paper (M\'enard et al. 2013, in preparation) showing that lowering the
dust temperature alone makes the line too low (i.e.,  does not change
the ratio properly) and that direct illumination of the disk rim
probably needs to be taken carefully into account  as well.
  
\subsection{Multiplicity}
\label{sec:mult}

The presence of stellar companions has potentially dramatic consequences
on the properties of circumstellar disks through the tidal forces 
exerted on them. For instance, it has been established that T Tauri
stars in close binaries ($\lesssim$50\,AU) are much less likely to host
disks  \citep{Cieza09, Kraus12}. Furthermore, there is evidence that
disks in these systems are much less massive than those in wider
binaries or around single stars (Harris et al. 2012, but see Duch\^ene
2010). It is thus natural to explore the connection between multiplicity
and our continuum and line observations. In particular, our survey
allows one to address this question for the gaseous component of
protoplanetary disks. To this end, we have compiled the multiplicity
properties of each system in our sample (Kenyon, Gomez \& Whitney 2008;
Kraus et al. 2011; White et al. 2013, in preparation). For high-order
multiple systems, we first evaluated which component hosts the disk
dominating the far-infrared emission, typically from high-resolution
sub-millimeter mapping \citep[e.g.,][]{Harris12}, and assigned the
separation of the system to the closest companion around that component 
but excluding  spectroscopic companions.

\begin{figure}
\centering
\includegraphics[width=6in]{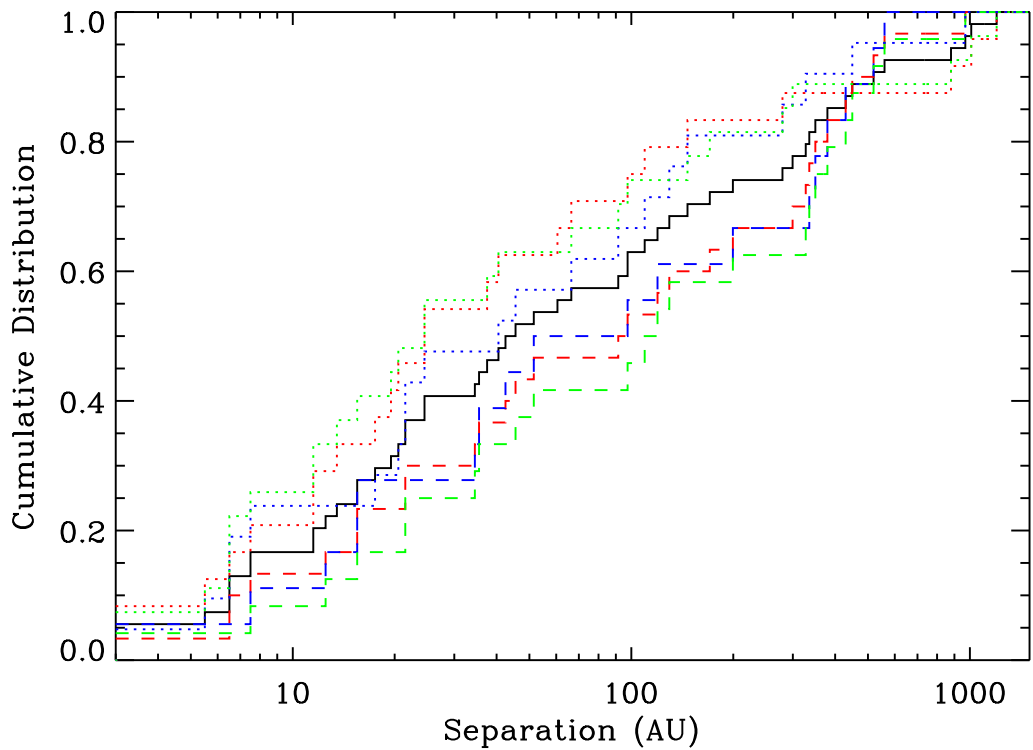}
\figcaption{Cumulative distribution of separations for binaries systems
in our sample. The black histogram represents all multiple systems in
our survey. Dashed and dotted histograms represent subsamples of object
which are respectively detected and non-detected in 70 $\mu$m (red) and
850 $\mu$m (green) continuum and [\ion{O}{1}] 63 $\mu$m (blue). Tight
binaries (below $\sim$50 AU) are more likely to be undetected in all
three types of observations than wider systems.\label{fig-cum}}
\end{figure}

\begin{figure}
\centering
\includegraphics[width=6in]{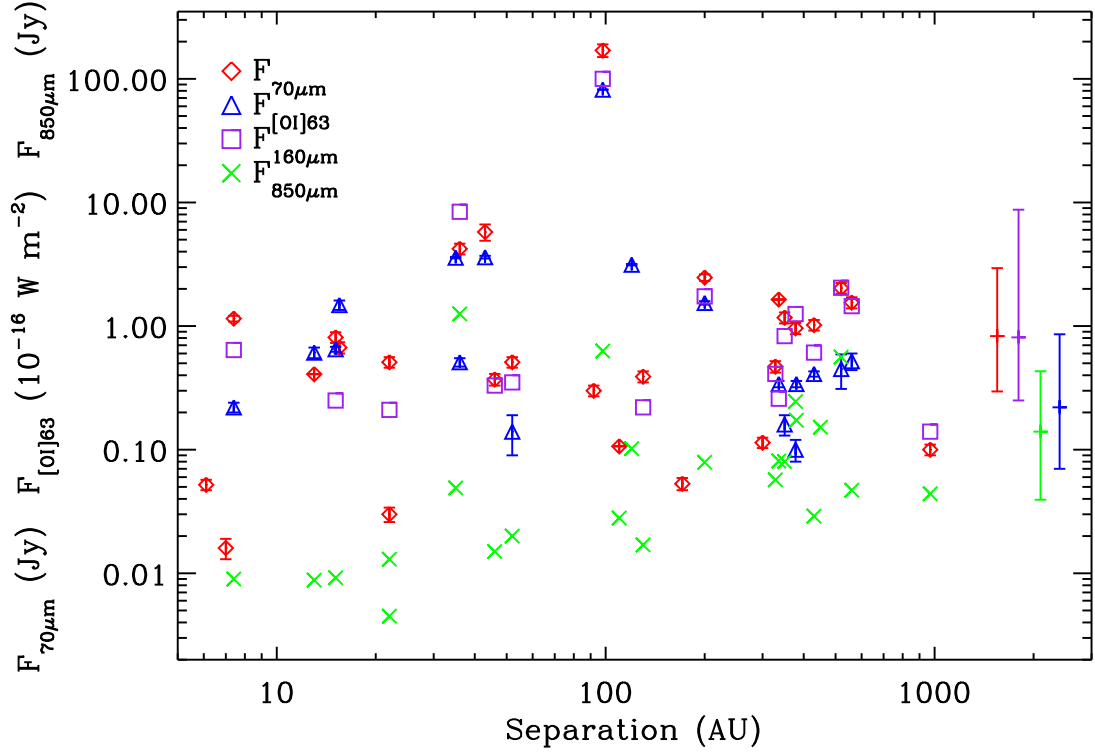}
\figcaption{Far-infrared and sub-millimeter continuum and line fluxes
for all multiple systems in our sample as a function of the system's
(projected) separation. For high-order multiple systems, we assigned to
the system the separation to the closest companion to the disk-bearing
component of the system to the exclusion of spectroscopic companions. We
plot here the continuum fluxes at 70 $\mu$m (PACS, red diamonds), 160 $\mu$m (PACS, purple squares), and 850
$\mu$m \citep[green squares]{Andrews05}, and the [\ion{O}{1}] 63 $\mu$m
line flux (PACS, blue triangles). Uncertainties displayed on PACS
datapoint do not include calibration uncertainties, whereas a typical
uncertainty of 10--20\% applies to all sub-millimeter points. Notice how
binaries tighter than 50--100\,AU have systematically reduced
sub-millimeter fluxes whereas their far infrared continuum and line
fluxes span the same range as those of wider systems. The ``error bars'' show the
median and the 34 percentile on each side (e.g.  one sigma range) for single
stars in our sample. 
\label{fig-sep}}
\end{figure}

The detection rate of multiple systems is lower than that of single
stars for both continuum and line emission. The continuum detection rate is
32/56 for binaries and  36/55 for singles with a significance level of 90\%. 
the line detection rate is 20/41 for binaries and
23/34 for singles with a significance level of 95\%.  Not surprisingly, we
find that systems tighter than 50--100\,AU are less likely to be
detected at far-infrared continuum or [OI] 63 $\mu$m emission than wider
systems (Figure~\ref{fig-cum}), similar to conclusions reached from
near- and mid-infrared as well as sub-millimeter wavelengths. Taken at
face value, this suggests that the gas and dust components of disks are
similarly affected by the presence of a companion. We note, however,
that the detailed interpretation of these conclusions is complicated by
the biased nature of the GASPS sample, which over-represents
disk-bearing systems, especially for spectroscopic observations.
Nonetheless, an intriguing finding is that neither the 70 $\mu$m nor the
[\ion{O}{1}] 63 $\mu$m line flux show a correlation with the binary
separation in systems for which they are detected  (Figure~\ref{fig-sep}). This is opposite the
results of sub-millimeter surveys  \citep[e.g.,][]{Harris12} but in line
with near- and mid-infrared studies \citep{Cieza09}, further suggesting
that both the continuum and line emission arise from the inner most
regions of the disk, which are not truncated by stellar companions
located 10\,AU or more from the disk-bearing star.

\subsection{Class III objects (weak-line T Tauri stars)}
\label{sec:ClassIII}

We do not detect line emission in any of the 15 weak-line T Tauri stars,
which we observed in [\ion{O}{1}] 63 $\mu$m (Table~\ref{tbl-lineflux}).
In continuum we observed 44 Class III objects and detected only three of
them: FW~Tau, V397 Aur, and V819 Tau. All three have 70 $\mu$m flux
densities far below our [\ion{O}{1}] 63 $\mu$m detection threshold, if
they follow the correlation found for ``disk-only'' sources, and would
therefore require unrealistically long integration times to reach the
expected signal level we predict from CTTS. V819 Tau (spectral type K7)
was classified as Class II by \citet{Luhman10} and Class III by
\citet{Rebull10}. It was detected at 70 $\mu$m by \citet{Cieza13}, who
did detailed SED modeling showing that the star has no excess in the
IRAC bands but a clear excess at 24 $\mu$m and 70 $\mu$m. Their modeling
suggests that V819 Tau has a warm debris disk. We detected  V819 Tau at
both 70 $\mu$m and 100 $\mu$m, because we have longer exposure times
than \citeauthor{Cieza13}. We see no trace of emission at 160 $\mu$m.
Both FW~Tau  and V397 Tau  were detected in all bands observed. Whether
these disks are at the end of their primordial phase  or whether the
stars have already reached the debris disk stage is impossible to judge
without detailed modeling.

 \subsection{Comments on individual objects}
\label{sec:individual}

Here we highlight a few stars, which previously were not known to
power jets or which appear to have unusual characteristics.\\

{\bf AA Tau}, spectral type K7 - M0 V, is a well-studied CTTS, which has
long been believed to power an optical jet
\citep{Hirth97,Bouvier99,Bouvier03,Bouvier07}. This was recently
confirmed by \citet{Cox13} who found  a poorly collimated, faint jet
extending to 21\arcsec\ from the star. The [\ion{O}{1}] $\lambda$6300
\AA\ line in AA Tau is narrow and slightly blue-shifted
\citep{Hirth97,Hartigan95}.  The  \ion{He}{1} $\lambda$10830 \AA\ line
is likewise faint , narrow and slightly blue-shifted \citep{Edwards06},
whereas it is expected to be deep, broad and blue-shifted for an outflow
source (c.f. discussion on DL Tau below).

\begin{figure}
\centering
\includegraphics[width=4in]{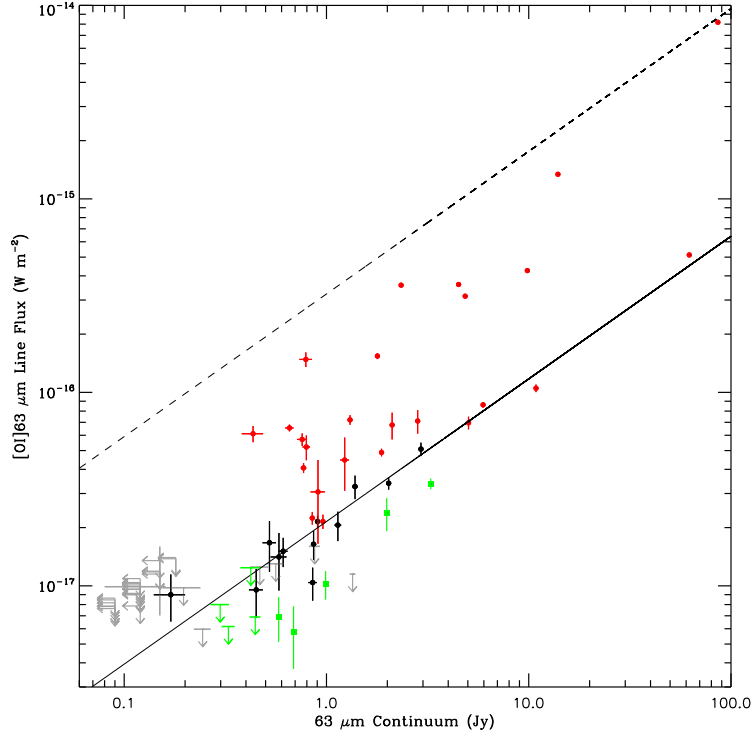}
\figcaption{ [\ion{O}{1}] 63 $\mu$m line emission versus 63 $\mu$m continuum emission for
the GASPS Taurus sample.  Non-outflow sources are plotted in black, outflow sources are
plotted in red, and transitional disks are plotted in green.  Sources for which there
is  less than 3$\sigma$ detection in line flux or continuum are plotted in grey as
3$\sigma$ upper limits.  None of the sources plotted in grey are known to drive outflows.
There is a tight correlation (lower black line) between  [\ion{O}{1}] line emission and
63 $\mu$m continuum flux for non-outflow sources (sources with less than 3$\sigma$
detection in line or continuum flux (the known outflow sources and the transitional disks
were omitted from the line fit), suggesting that the emission originates from the same
part of the disk. The  [\ion{O}{1}] 63 $\mu$m line emission in outflow sources is
dominated by the outflow, and can be up to twenty times stronger (upper dashed line) than
the emission from the disk.   
\label{corr}}
\end{figure}

\citet{Baldovin11} searched for mid-IR emission lines in a sample of 64
PMS stars in Taurus and detected [\ion{Ne}{2}] 12.81 $\mu$m emission in
18 objects including AA Tau. They found in general that  the luminosity
of the [\ion{Ne}{2}] line was stronger for sources driving jets than
that for those without known jets, although not  for AA Tau.
\cite{Najita09}, who analyzed  high-resolution [\ion{Ne}{2}] emission
line profiles from AA Tau found that they could be explained by
originating in  the disk, although they could not rule out contribution
from a jet. Our [\ion{O}{1}] 63 $\mu$m observations, however, show that
[\ion{O}{1}] line is consistent with a disk origin (Figure~\ref{corr}),
i.e., with little or no excess emission from the jet. However, since it
has been shown to drive a faint jet,  we classify it as a jet/outflow
source. The inner disk of AA Tau is rich in organic molecules
\citep{Carr08}. The o-H$_2$O line at 63.32 $\mu$m has also been detected
for this source \citep{Riviere-Marichalar12}.\\

{\bf CoKu Tau/4} is a M1.5 T Tauri star with  weak H$\alpha$ emission
\citep{Cohen79,Kenyon98} and it therefore has a low accretion rate. It
was indentified by \citet{Forrest04} as a transitional disk based on
{\it Spitzer} IRS observations, which revealed that it has no infrared
excess shortward of 8 $\mu$m, but a large excess at 20 - 30 $\mu$m 
indicating that the region within 10 AU  have been largely cleared of
dust. \citet{Ireland08} carried out aperture-masking interferometry and
adaptive optics imaging of the star and show that CoKu Tau/4 is  a
near-equal binary star of projected separation $\sim$ 8 AU (53 mas). The
binary has therefore cleared the inner disk. Both stars are of spectral
type M1 - M2  and are $\sim$ 3 - 4~Myr old \citep{Ireland08}. It is therefore
not a transitional disk.

\begin{figure}
\includegraphics[ scale=0.9,angle=-90]{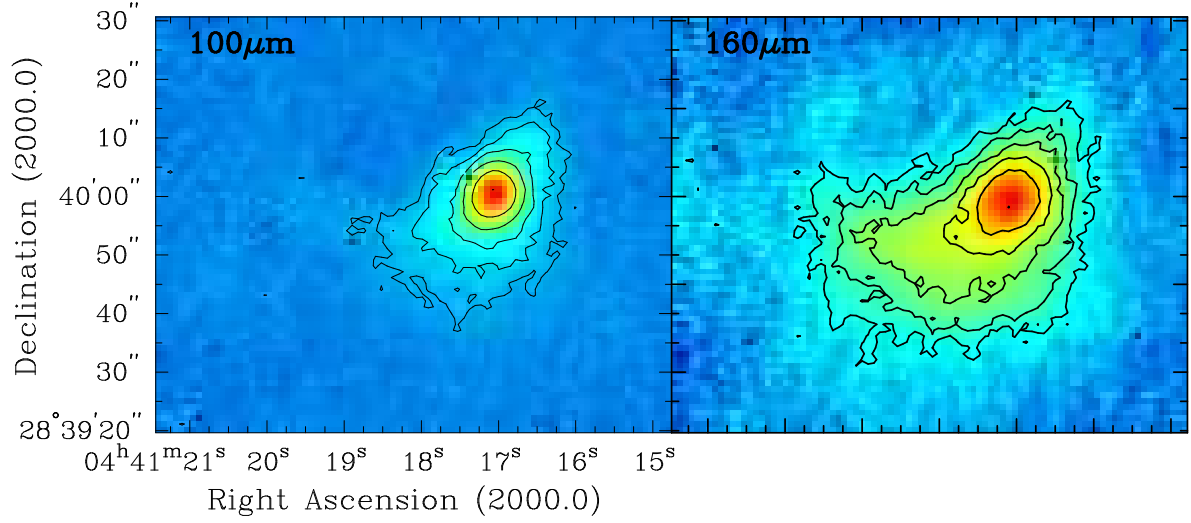}
\figcaption[]{
\label{fig-CoKu}
False color images of CoKu Tau/4 overlaid with logarithmic contours.  CoKu Tau/4 is
surrounded by an extensive far infrared nebulosity at 160 $\mu$m, which is much fainter at 100
$\mu$m.}
\end{figure}

\begin{figure}
\includegraphics[width=6in]{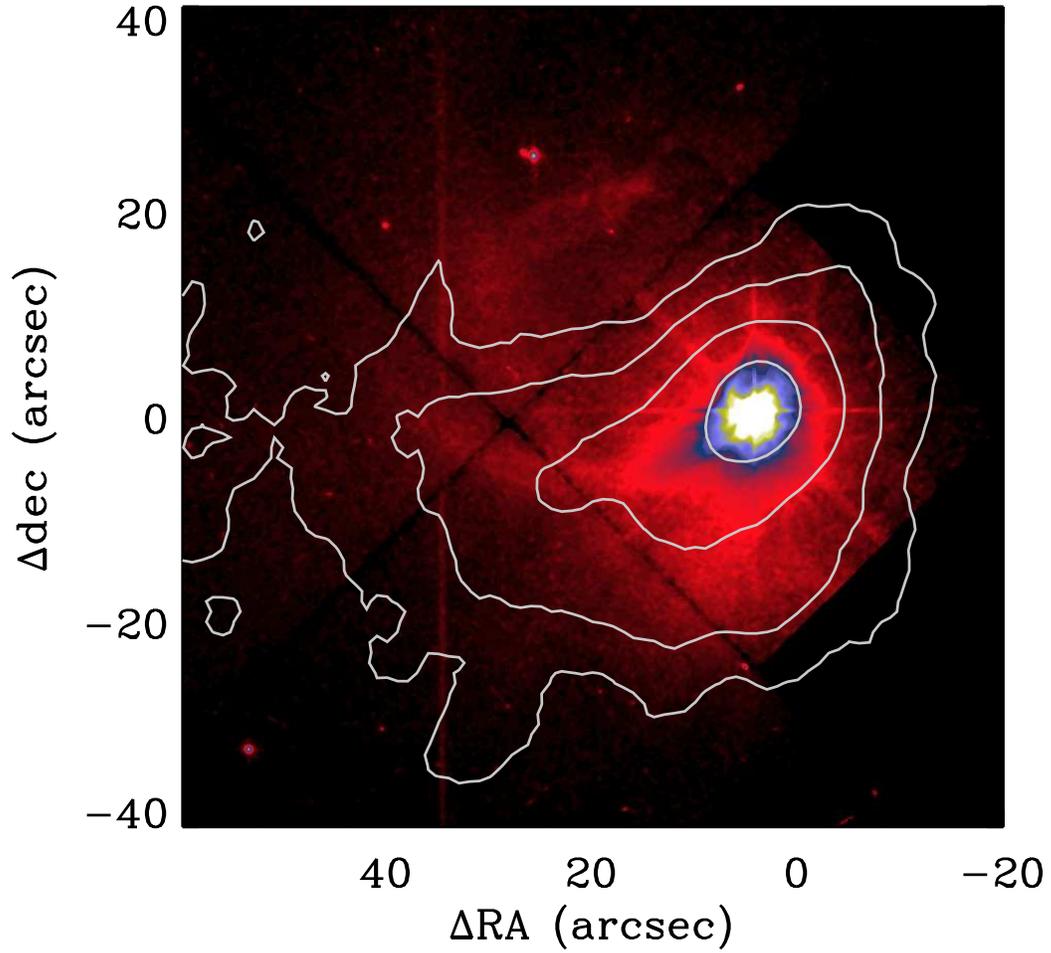}
\figcaption[]{
\label{GD}
WFPC 2 image (0.8 $\mu$m)
of the reflection nebulosity illuminated by  CoKu Tau/4.  The PACS 160 $\mu$m image
(contours) closely follows the nebulosity. }
\end{figure}

CoKu Tau/4 is a strong far-infrared source (Table~\ref{tbl-cont}) and
one of the few stars in our sample which is associated with a
far-infrared nebulosity (Figure \ref{fig-CoKu}). At 70 $\mu$m the star
appears point-like, but at 100 $\mu$m nebulosity can be seen to the
south and south east of the star. At 160 $\mu$m the emission extends
over more than an arcminute  (Figure~\ref{fig-CoKu}). The far infrared
nebulosity follows closely the reflection nebulosity seen with WFPC 2
camera on the Hubble Space Telescope \footnote{The WFPC 2 image for CoKu
Tau 4 was obtained as part of HST programme GO 9160, PI D.L.
Padgett.}(Figure \ref{GD}). The [\ion{O}{1}] flux is consistent with a
disk origin (Figure~\ref{corr}). \\

{\bf DE Tau} is a young CTTS of spectral type M2  \citep{Kenyon95}. It
is not known to drive an outflow. It stands out in our sample as the
only non-outflow star with a $3\sigma$ detection of the [\ion{O}{1}] 145
$\mu$m line and only  a marginal ($\sim2\sigma$) detection of the 
[\ion{O}{1}]  63 $\mu$m line. For DE Tau the upper limit of the line
ratio of [\ion{O}{1}]  63 to  [\ion{O}{1}] 145 $\mu$m is on the order of
unity, suggesting that both transitions are optically thick and
originate from the same part of the disk. An isothermal greybody fit to
the millimeter and far infrared SED requires the dust to be optically
thick at millimeter wavelengths, i.e., a dust emissivity index of 0,
suggesting that DE Tau could be surrounded by a compact optically thick
disk.\\

{\bf DG Tau \& DG Tau B} both power optical jets
\citep{Mundt83,Mundt87,Kepner93,Eisloffel98}. DG Tau is a young, heavily
accreting star with a spectral type of  K7-M0 \citep{Gullbring98},
obscured by a visual extinction of 3.2 - 5.4 mag \citep{Fischer11}. It
also powers  a low velocity, wide-angled bipolar molecular outflow (G. 
Sandell, 2013, private communication).  The disk has been imaged with
high spatial resolution at mm-wavelengths and is seen relatively face on
with an inclination of $\sim$ 30 \degr\ \citep{Isella10,Guilloteau11}.
In this case we find the [\ion{O}{1}] 63 $\mu$m emission to be
completely dominated by the outflow but only marginally extended.
\citet{Podio13} spectrally resolved the   [\ion{C}{2}]  158 $\mu$m line
with HIFI, which shows that the   [\ion{C}{2}]  emission originates in
the blue-shifted jet, and not in the disk.

DG Tau B is a deeply embedded Class I  source. The spectral type is
unknown, but modeling of the SED suggests that it is low mass ($\leq$
0.3 \Msun), i.e. mid M. It powers a  well collimated bipolar molecular
outflow \citep{Mitchell97}. Compared to DG Tau the  [\ion{O}{1}] 63
$\mu$m emission is clearly extended and much stronger in the red-shifted
outflow lobe than on the star \citep{Podio12}. At 70 $\mu$m the
continuum emission is unresolved on both stars, but at 100 $\mu$m and
especially 160 $\mu$m one can see faint emission from the surrounding
dark cloud, possibly from dust being heated and compressed by the
outflows that are powered by DG Tau and DG Tau B. The PACS line
spectroscopy from both stars is discussed in detail by
\citet{Podio12}.\\

{\bf DL Tau}, a CTTS of spectral type K7 V, is classified as an
outflow/jet source based only on the broad [\ion{O}{1}]$\lambda$6300
\AA\  and [\ion{S}{2}] $\lambda$6371 \AA\ \citep{Hartigan95}. The
[\ion{O}{1}] $\lambda$6300 \AA\  line shows a high velocity component,
which is blue-shifted to $\sim$ -300 km~s$^{-1}$, while the [\ion{S}{2}]
$\lambda$6371 \AA\  goes to -250 km~s$^{-1}$, and has no low velocity
component at all. DL Tau has deep and broad blue-shifted  \ion{He}{1}
$\lambda$10830 \AA\ \citep{Edwards03,Edwards06}, which \citet{Kwan11}
show originates in an outflow rather than from accretion. C. Grady
(2013, private communication) reports that a jet as well as two distant
HH objects have recently been discovered in DL Tau. The disk is seen
relatively face-on with an inclination {\it i} =  38 $\pm$ 2$^{\circ}$
\citep{Guilloteau11}. We find the [\ion{O}{1}] $\lambda$63 $\mu$m
emission to be in the outflow regime (Figure~\ref{corr}), which confirms
that DL Tau indeed powers an outflow.\\

{\bf GG Tau} is a hierarchical quadruple system with a 10\arcsec\
separation between the two binaries \citep{Leinert91}. GG Tau was one of
the brightest mm-sources in the survey by \citet{Beckwith90} suggesting
that it must be surrounded  by a massive circumstellar disk. Millimeter
interferometry \citep{Simon92,Guilloteau99}  shows that GG Tau Aa/Ab is
surrounded by a large circumbinary disk with an outer  radius of 2 -
4\arcsec. GG Tau Aa has a spectral type of K7, while the secondary,
GG Tau Ab, has a spectral type of M0.5 \citep{White99}. The inner part of
the ring/disk has been resolved in high-resolution scattered light images 
from 0.5 to 4 $\mu$m   \citep{Roddier96,McCabe02,Duchene04,Krist05} and found to be elliptical with a semimajor axis
of 1\farcs5. 

\begin{figure}
\includegraphics[ scale=0.67,angle=-90]{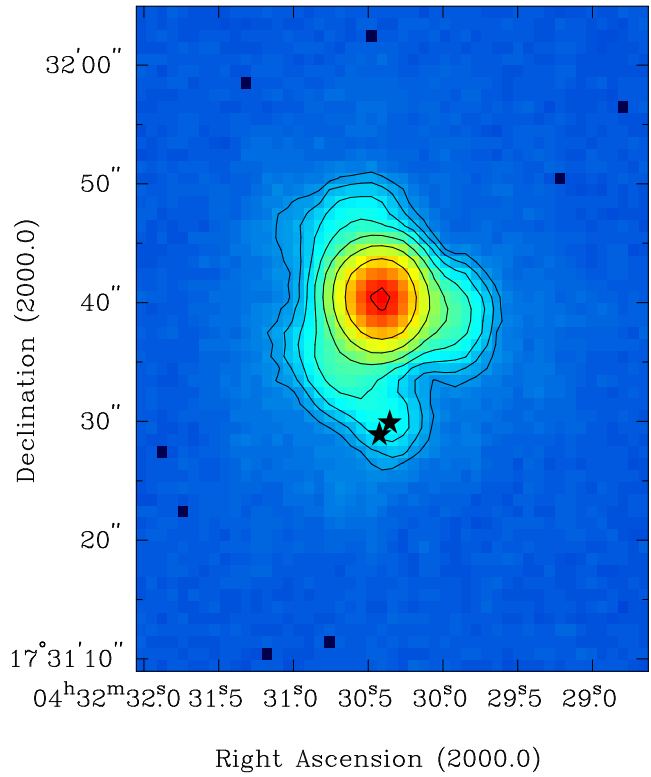}
\figcaption[]{
\label{fig-ggtau}
 Logarithmically stretched 70 $\mu$m image of GG Tau showing the trifoil structure of the PSF.  The additional spur to the south
coincides with the southern binary, GG Tau Bab, which is shown with star symbols. A PSF subtraction (not shown) confirms the detection of GG Tau  B.
}
\end{figure}

The secondary pair, GG Tau Ba and Bb is also a binary and separated 
from the primary by 10\farcs4 at a PA of 185\degr{} \citep{Kraus09}.  GG Tau Ba
has a spectral type of M5 and Bb M7 \citep{White99}. We detect faint emission from the system at 70 $\mu$m (Figure~\ref{fig-ggtau}).\\

\begin{figure}
\includegraphics[ scale=0.8,angle=0]{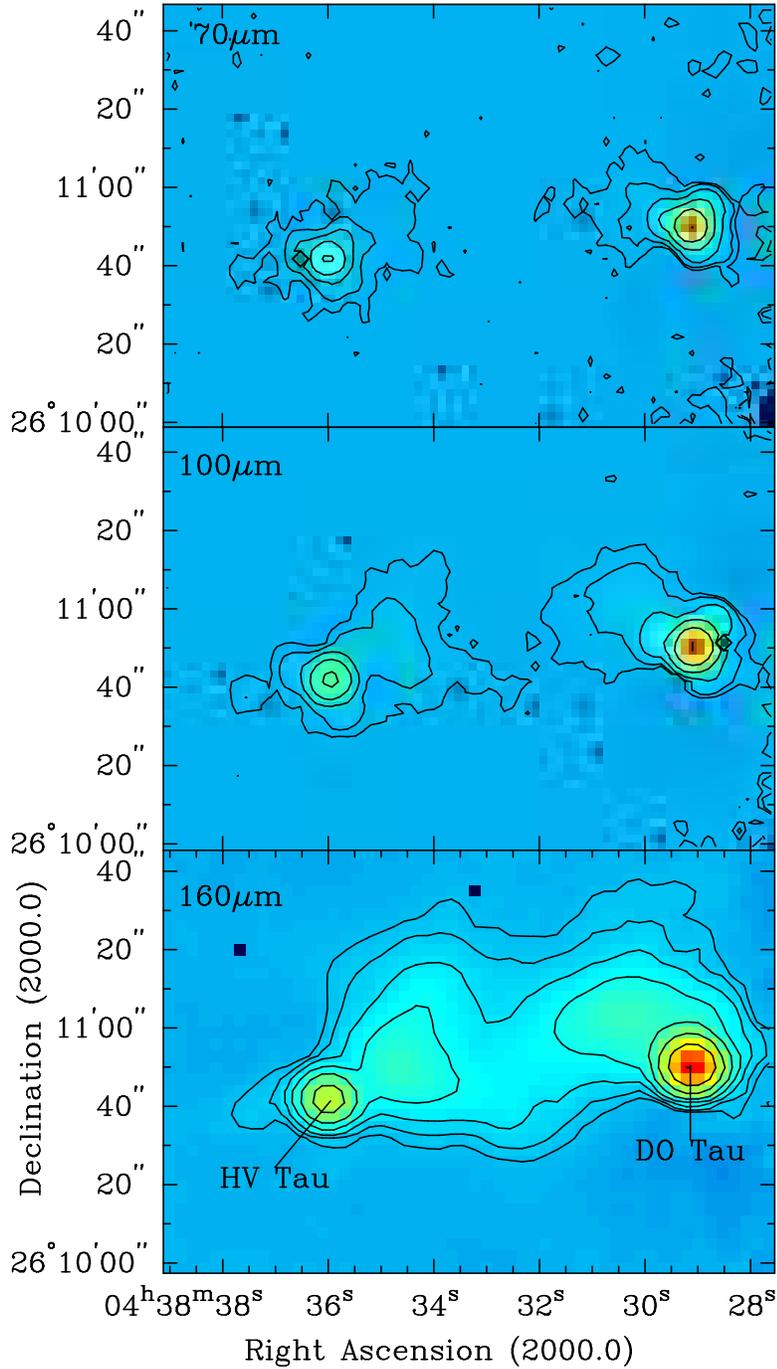}
\figcaption[]{
\label{fig-hvtau}
False color images of HV Tau and DO Tau  at 70  $\mu$m, 100 $\mu$m, and 160 $\mu$m
overlaid with logarithmic contours.
DO Tau is close to the edge of the image, which results in excess noise east
of the star. Both stars appear to be associated with extended emission,
which is barely visible at 70 $\mu$m. At 160  $\mu$m the emission is very
extended. 
}
\end{figure}

{\bf HV Tau and DO Tau} are  two CTTSs associated with far infrared
nebulosities. The separation between the two is $\sim$ 90.8 \arcsec\ and
both were imaged simultaneously with the PACS imager, although DO Tau is
close to the edge of the image.  Both stars are associated with far
infrared nebulosities, which become visible at 100 $\mu$m
(Figure~\ref{fig-hvtau}). At 160  $\mu$m the emission is very extended
and connect both stars in a common envelope.

DO Tau is a G star associated with an arc-like nebula, which is aligned
with the nebulosity we see at 100 and 160 $\mu$m \citep{McGroarty04}. It
drives a bipolar jet with the redshifted jet at a PA of 70\degr\
\citep{Hirth94}, approximately aligned with the arc-like nebulosity we
see at 100 $\mu$m. \citet{McGroarty04} found three Herbig-Haro (HH) objects at PAs of
74 - 78\degr\ northeast of DO Tau. The farthest one, HH 831\,B is
11\arcmin\ from DO Tau.

HV Tau is a close binary \citep{Simon96}, which show no infrared excess
\citep{Woitas98}.  About 4\arcsec\ to the north east is a third star, HV
Tau C, which has an edge on disk \citep{Monin00,Terada07}. The spectral
type of HV Tau C is K6 \citep{White04}. This is the star we see in the
far infrared, and which powers a bipolar jet
\citep{Stapelfeldt03,McGroarty04,Duchene10a} with the PA of the blue-shifted jet
being $\sim$ 25\degr\ \citep{McGroarty04}. In this case the V-shaped 100
$\mu$m emission is aligned with a diffuse reflection nebula 
west-northwest of the star \citep{Martin94}, which appears to be
approximately aligned with the disk plane, and orthogonal to the jet.

We find that both stars have strong [\ion{O}{1}] 63 $\mu$m emission, as
expected from stars powering jets.\\

{\bf HL Tau and XZ Tau} are  two  of the first T Tauri stars found to
power  well-collimated optical jets
\citep{Mundt83,Mundt90,Movsessian07}. HL Tau is a Class I object of
spectral type K7 \citep{White01}. It has very strong high velocity
blue-shifted  [\ion{O}{1}] $\lambda$ 6300 \AA\ emission
\citep{Edwards87} and it is one of the brightest T Tauri stars at 1.3 mm
\citep{Beckwith90}.  The disk has an inclination angle of $\sim$
42\degr\ \citep{Lay97}. The  [\ion{O}{1}] 63 $\mu$m emission is
definitely in the outflow regime (Figure~\ref{corr}), but it does not
have a particularly strong excess. XZ Tau, which has a lower accretion
rate, has a much larger excess in   [\ion{O}{1}] 63 $\mu$m due to the
outflow.\\

{\bf RY Tau} is classified as a F8 - G1 type star
\citep{Mora01,Calvet04}, although \citet{Kenyon95} assigned it a
spectral type as late as K1. It is a well known outflow source, which
drives a well-collimated jet extending out at least 31\arcsec\ from the
star and with a counter-jet extending out to at least 3.5\arcmin\ in the
opposite direction \citep{St-Onge08}. The star has relatively strong
free-free emission \citep{Rodmann06}, which is consistent with a thermal
wind. The spectro-imaging observations by \citet{Agra-Amboage09} detect
the blue-shifted [\ion{O}{1}] $\lambda$6300 \AA\  within 2\arcsec\ of
the star, consistent with the large scale jet seen by \citet{St-Onge08}.
\citet{Agra-Amboage09} estimate the inclination of the jet to be between
45\degr\ - 77\degr. \citet{Isella10} resolved RY Tau at 1.3 mm with
millimeter interferometry and found that it has a large central cavity 
(28 AU) similar to what is seen in transitional disks. Alternatively the
dust grains in the inner disk have grown to centimeter sizes lowering
the dust opacity. \citet{Espaillat11} found that they could fit the SED
of RY Tau with an 18 AU gap which contains some optically thin dust
consistent with the cavity observed by \citet{Isella10}.

RY Tau is also peculiar in other respects. \citet{Lommen10} find a
correlation between the strength of the 10-$\mu$m silicate feature and
the slope of the mm-emission measured between 3 and 1 mm, except for RY
Tau, which in their data set is an extreme outlier, although this is not
entirely true. Our analysis, based on all published
millimeter/sub-millimeter data, shows that the millimeter slope is
rather normal, therefore in that respect RY Tau is not that unusual. The
strong silicate emission, however, requires an abundance of small
grains. What is unique though is that a  transitional disk star would
drive a well collimated optical jet, have free-free emission and  still
be rather strongly accreting
\citep[$\dot{M}$$\sim$6--9$\times$10$^{-8}~{\rm M}_{\odot}$
yr$^{-1}$;][] {Agra-Amboage09}.

Our observations shows no excess from the jet  in the  [\ion{O}{1}] 63
$\mu$m line. The line  intensity is completely consistent with a disk
origin (Figure~\ref{corr}).  This could mean either  that there is no
[\ion{O}{1}] 63 $\mu$m emission from the outflow or that the emission
from the disk is anomalously low, perhaps due to the central cavity.\\

{\bf StHA 34} (HBC 425) is a spectroscopic binary with components of
nearly equal luminosity and temperature (both M3) and broad, strong
H$\alpha$ emission characteristic of a CTTS \citep{White05}. However,
neither component of the binary shows \ion{Li}{1} $\lambda$6708
absorption suggesting a very long-lived accretion disk. Comparison with
PMS evolutionary models give an isochronal age of 8 $\pm$ 3 Myr, which
is much younger than the predicted lithium-depletion timescale of
$\sim$25 Myr. \citet{Hartmann05}, who observed StHA 34 with IRS on {\it
Spitzer} modeled the infrared SED with three components: an inner disk
wall, an optically thin inner disk, and an outer disk, i.e., similar to
the SED of a transitional disk. They favored an age of 25 Myr, which
would require StHA34 to  be at a distance of $\sim$100 pc, and therefore
not a member of the Taurus association.

\citet{Dahm11} found a low mass companion $\sim$1\farcs23 southeast of
the primary pair, which has strong \ion{Li}{1} $\lambda$6708 absorption.
STHA 34 C has a spectral type of M5.5. Comparison with PMS evolutionary
tracks imply a mass of $\sim$0.09 \Msun\ and an age of 8 - 10 Myr
assuming the nominal distance to Taurus (140 pc). It therefore seems
more likely that StHA 34 is lithium depleted and not surrounded with a
25 Myr old, still accreting circumbinary disk.

We detected StHA 34 in continuum at 70, 100, and 160 $\mu$m 
(Table~\ref{tbl-cont}). The emission is rather faint, which is
consistent with an evolved, possibly transitional disk.

{\bf T Tauri} is an exceptional member of the class of pre-main sequence
stars that share its name. It is a hierarchical triple system with the
optical star T Tau N having a K-type spectrum and large optical
variations on timescales of months to years \citep{Beck01}, and the
southern deeply embedded binary system T Tau S, which has a projected
separation of 0\farcs05  \citep{Koresko00}. T Tau N is known to power a
bright optical jet \citep{Buhrke86}.  T Tau S is surrounded by a massive
accretion disk and powers a spectacular jet \citep{Reipurth97}.

ISO found T Tauri to have the richest line mid/FIR  emission line
spectrum of any PMS star \citep{Lorenzetti05}.  T Tauri is very bright
in the far infrared (Table 2), and the emission is extended ($\sim$
4\arcsec{}) in all PACS bands. We see faint emission from the Hind's
nebula at 70 $\mu$m, which is much stronger at 100 and 160 $\mu$m. 
At 160 $\mu$m we also start to pick up emission from the surrounding
cloud.

Since T Tauri is a triple system which drives at least two optical jets, it is not clear
where the emission originates.  It could arise in the disk, in the outflow, or in the hot
surrounding envelope \citep{Podio12}.  We have therefore excluded it from any correlation analysis.\\

{\bf V773 Tau} is a compact  quadruple system with at least four stars
within  0\farcs3 \citep{Boden12}. V773 Tau A is a spectroscopic binary
with spectral types of  K2 and K5 \citep{Welty95}.  It is a weak-line T
Tauri star (WTTS) and one of the strongest radio stars in Taurus 
\citep{ONeal90,Massi08}. The radio emission is non-thermal and strongly
variable \citep{Dutrey96}. The variability is due to interacting coronae, which causes
strong flaring at periastron \citep{Massi08}. V773 Tau A has no near
infrared excess, and is therefore unlikely to be surrounded by a
circumbinary disk. V773 Tau B is a CTTS, which varies by more than 0.5 mag in the
visible and has a spectral type of K7 - M0.5 \citep{White01,Duchene03,
Boden12}. It has some near-IR excess longwards of 2 $\mu$m and maybe
surrounded by a compact  disk. V773 Tau C is
extremely red and  almost certainly surrounded by a circumstellar disk
\citep{Duchene03}. Since V773 Tau was detected at 1.3 mm and 850 $\mu$m
\citep{Osterloh95,Andrews05} it is clear that the system must have a
circumstellar disk. Furthermore we easily detected it at
both 60 and 160 $\mu$m (Table 2). Whether the dust emission is  dominated
by V773 Tau C or whether V773 Tau B also contributes  requires
observations with sub-arcsecond
imaging resolution with good sensitivity, as can be provided by ALMA.

V773 Tau is not known to drive a jet, yet the [\ion{O}{1}] 63
$\mu$m flux density clearly puts it in the outflow regime
(Figure~\ref{corr}). \citet{Cabrit90} did detect [\ion{O}{1}] $\lambda$
6300  \AA\  and [\ion{S}{2}] $\lambda$ 6731 \AA\ emission but did not
publish a spectrum. It is not clear whether the emission comes from a
jet, especially since the velocity offsets were small. However, FM Tau,
an M0 star, is 38\arcsec\ to the north-northwest of V773 Tau, and CW
Tau,  a K3 star, 92\arcsec\ to southeast.  The latter is a young Class
II type CTTS, which powers a bipolar optical jet and excites several HH
objects \citep{Dougados00,McGroarty04}.  Not all the HH objects
discovered by \citet{McGroarty04}  can be explained by the outflow from
CW Tau. In particular,  HH 828, which is just north of FM Tau and V773 Tau,
is not  aligned with the CW Tau jet, and therefore more likely to be
powered either by FM Tau or V773 Tau. The [\ion{O}{1}] $\lambda$ 6300
\AA\  line in FM Tau is slightly blue-shifted and does not show much
high velocity emission \citep{Hartigan95}, and is therefore an unlikely source. 
V773 Tau may well be exciting HH 828. In our
analysis we therefore assume that V773 Tau is an outflow source,
although this still needs to be confirmed.

\section{Summary and Conclusions}
\label{sec:summary}

 In this paper we present the results of Herschel/PACS photometric and
spectroscopic observations of a large sample of T Tauri stars in the
Taurus/Auriga star-forming regions as part of the GASPS survey. We
observed 82 fields in photometry and 75 in spectroscopy, for a total
sample of about 120 objects, mostly Class II/III objects (CTTS/WTTS)
as well as a handful of embedded Class I sources and transitional
disks.

In photometry, we detected 50 known members of the association, apart
for 3 Class III sources (FW~Tau, StHA 34, and V819 Tau), all of them
are Class II or Class I sources. Most objects are spatially unresolved in the
continuum. Exceptions include T Tau, the multiple systems UZ Tau and
GG Tau and HD 283759 (but the latter is not a member of the association).

Even though our spectroscopic observations include a slew of other
molecular transitions, we only discuss here the results for
[\ion{O}{1}] 63 $\mu$m and 145 $\mu$m, as well as [\ion{C}{2}] 158
$\mu$m. Results on the o-H$_2$O line at 63.32 $\mu$m were presented in
\citet{Riviere-Marichalar12} while other transitions, typically
detected in strong outflow sources and sometimes spatially resolved,
were discussed in \citet{Podio12} for a subset of our sample.

The [\ion{O}{1}] 63 $\mu$m line, spatially and spectrally unresolved
in most sources, is by far the strongest line in our sample and often
the only line detected, especially for fainter sources. With a typical
line sensitivity of $\sim$ 1$\times10^{-17}$ W m$^{-2}$, we obtained
$>3\sigma$ detections in 43 out of 75 targets; no Class III sources
were detected. Marginal (2--3$\sigma$ level) detections are presented
for another 10 sources. The [\ion{O}{1}] 145 $\mu$m line is 10--25
times fainter than the [\ion{O}{1}] 63 $\mu$m line and detected only
in 17 sources. The detection rate for [\ion{C}{2}] 158 $\mu$m line is
even lower (10 targets) and detections are restricted to jet/outflow
sources.

We found no correlation between [\ion{O}{1}] 63 $\mu$m line intensity
and disk mass, suggesting that the line is either optically thick or
is not a valuable direct probe the mass of the overall gas
reservoir. However, we found a tight correlation between the
[\ion{O}{1}] 63 $\mu$m line strength and the 63 $\mu$m continuum flux
density for ``disk-only'' objects, i..e, sources with no or only weak
outflow activity. Sources with clear evidence for outflow activity, on
the other hand, can show line strengths that are as much as 20 times
higher than disk-only sources, suggesting that the line emission is
dominated by the jet. Some outflow sources (6 out of 24), however, follow the disk correlation and show little or no excess due to the outflow. Among disk sources, transitional disks also
stand out, being fainter in [\ion{O}{1}] than classical T
Tauri stars by a factor of 2--3 for comparable continuum flux
densities. 

The tight correlation between [\ion{O}{1}] line and 63 $\mu$m
continuum emissions in the disk-only sample suggests that they
originate from the same region of the disk. Simple disk models
indicate that the most of the far-infrared continuum emission arise
from the inner disk regions (within 5--50 AU at most, depending on the
disk outer radius). Based on two-component isothermal grey body fits
to the millimeter and far-infrared SEDs, we suggest that the
[\ion{O}{1}] 63 $\mu$m emission and the 63 $\mu$m continuum emission
originate from the inner disk and/or lower surface layers where the
gas and the dust is thermalized. This can also explain the weaker line
emission of transition disks,  since the dust in these systems is
substantially cooler. The absence of dependence of both the
[\ion{O}{1}] 63 $\mu$m line and far-infrared continuum emission
strength on binary separation for systems wider than 10\,AU further
supports the co-spatiality of their emitting regions and the fact that
they are limited to the innermost regions of the disk. 

The observations presented in this paper provide a rich dataset to
test thermo-chemical models which will provide a better understanding of the origin and
strength of the [\ion{O}{1}] 63 $\mu$m line in protoplanetary disks.

 \acknowledgements
This work made extensive use of the SIMBAD
Database, operated at CDS, Strasbourg,
France, and NASAÕs Astrophysics Data System Abstract
Service. We thank Uma Gorti for stimulating discussions. We thank the referee for
a very careful reading of the manuscript and for his/her constructive criticism.
F. M\'enard acknowledges financial support from the Milenium Nucleus P10-022-F, 
funded by the Chilean Government, and by the EU FP7-2011 programme, under Grant Agreement 284405.

\clearpage

\begin{figure}
\centering
\includegraphics[width=5in]{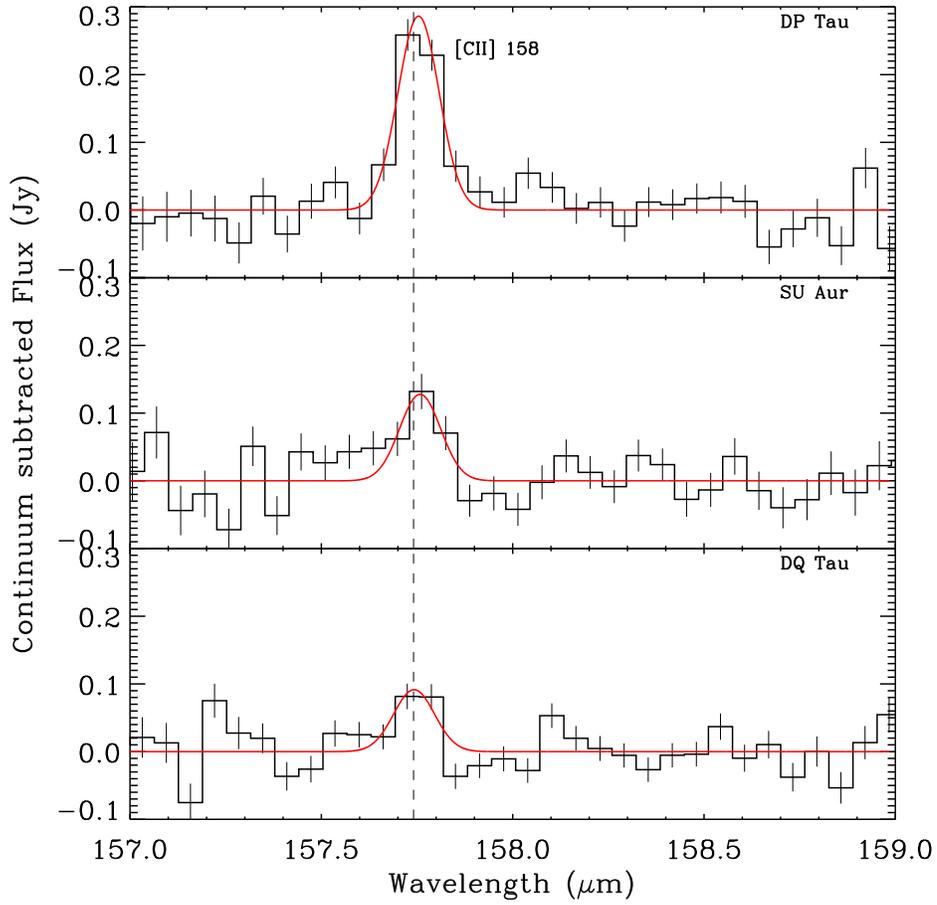}
\figcaption{ [\ion{C}{2}] 158  line detections for three of the targets in the Taurus
sample.  DQ Tau (bottom panel) is a non-detection. The line seen at $\lesssim$
2$\sigma$ is almost certainly spurious.     
\label{CII158}}
\end{figure}

\newpage

\begin{deluxetable}{lcrrrl}
\tabletypesize{\footnotesize}
\tablecaption{GASPS Taurus Observations:PACS Photometry \label{tbl-cont}}
\tablewidth{0pt}
\tablehead{
\colhead{Name} & \colhead{SED} &\colhead{70 $\mu$m}  & \colhead{100 $\mu$m}  & \colhead{160 $\mu$m} & \colhead{Comments}  \\
\colhead{} & \colhead{Class} & \colhead{[Jy]} & \colhead{[Jy]} & \colhead{[Jy]}  & \colhead{} }
\startdata
CIDA 2    & III   &  $<$ 0.009        & $<$ 0.009     & $<$ 0.022 &      Empty field\\
CIDA 10   & III   &   $<$ 0.008     &   \nodata  &  $<$ 0.028 &   Empty field\\
CoKu Tau 2 & I &  1.30 $\pm$ 0.20 & 1.32 $\pm$ 0.20 &  1.53 $\pm$ 0.30 & 34\arcsec\  West of HL Tau \\
CoKu Tau 4\tablenotemark{a}& II & 1.15 $\pm$ 0.06  & 0.85 $\pm$ 0.08    & 0.64 $\pm$ 0.10 &    \\
CW Tau\tablenotemark{b}  & II & 1.79 $\pm$ 0.18    & \nodata            & 1.82 $\pm$ 0.18 &  Same field as V773 Tau  \\
CX Tau\tablenotemark{c}  & T & 0.33 $\pm$ 0.03    & 0.25 $\pm$ 0.03    & 0.25 $\pm$ 0.03  &   S1 \\ 
CY Tau   & II & 0.22 $\pm$ 0.01    & 0.29 $\pm$ 0.03    & 0.35 $\pm$ 0.04   & S1 \\
DG Tau  & II & 23.8 $\pm$ 2.4 & 23.3 $\pm$ 2.3  & 18.30 $\pm$ 2.0 & Cloud emission, S2 \\
DG Tau B & I & 15.4 $\pm$ 1.5 & 18.0 $\pm$ 1.8  & 18.10 $\pm$ 2.0 & Same field as DG Tau \\
DH Tau A & II & 0.47 $\pm$ 0.05 & 0.48 $\pm$ 0.05 & 0.41 $\pm$ 0.04 & \\
DI Tau  AB & III & $<$ 0.01      & $<$ 0.01& $<$ 0.03 & near DH Tau \\ 
DK Tau A & II & 1.17 $\pm$ 0.12  & 1.07 $\pm$ 0.11 & 0.83 $\pm$ 0.08 & \\
DL Tau  & II  & 1.32 $\pm$ 0.13    & 1.54 $\pm$ 0.15    & 1.93 $\pm$ 0.19 &  S1 \\
DM Tau  & T & 0.78 $\pm$ 0.08 & 0.85 $\pm$ 0.08 & 0.78 $\pm$ 0.08 & \\
DO Tau\tablenotemark{d} & II  & 5.1\phantom{0} $\pm$ 0.80    & 4.7\phantom{0} $\pm$ 0.70    & 4.1\phantom{0}  $\pm$ 0.70 &    \\ 
DP Tau  & II & 0.67 $\pm$ 0.07    & 0.36 $\pm$ 0.04    & 0.20 $\pm$ 0.03  &  Cloud emission  \\
DQ Tau\tablenotemark{e} &II   & 1.37 $\pm$ 0.10    & 1.28 $\pm$ 0.12    & 1.02 $\pm$ 0.10 &   S1, field includes Haro 6-37 \\
DS Tau  &II  & 0.22 $\pm$ 0.02    & 0.33 $\pm$ 0.03    & 0.26 $\pm$ 0.03 &   \\  
FF Tau AB   &III &  $<$ 0.009       &  $<$ 0.010       &   $<$  0.028     &  S2 \\
FM Tau\tablenotemark{b} & II   & 0.47 $\pm$ 0.07    &  \nodata            & 0.24 $\pm$ 0.03 &    \\
FO Tau AB  & T & 0.51$\pm$ 0.05    & 0.39 $\pm$ 0.04    & 0.21 $\pm$ 0.03    & S1 \\
FP Tau\tablenotemark{c}  & II  &  0.33 $\pm$ 0.03    & 0.34 $\pm$ 0.06    & 0.38 $\pm$ 0.04  &   S1\\ 
FT Tau & II &  0.73 $\pm$ 0.07 & 0.95 $\pm$ 0.09 & 1.27 $\pm$ 0.19 & \\
FW Tau ABC & III &  0.030 $\pm$ 0.004 &  0.033 $\pm$ 0.004 & 0.070 $\pm$ 0.040 & S1, cloud emission \\
FX Tau  A & II &  0.39 $\pm$ 0.04 & 0.36 $\pm$ 0.04 & 0.22 $\pm$ 0.03 & \\ 
GG Tau Aab &II &  4.22 $\pm$ 0.42 & 6.56 $\pm$ 0.66 & 8.41 $\pm$ 0.84 & Extended \\
GG Tau Bab & II & 0.21 $\pm$ 0.02 & \nodata & \nodata & \\
GH Tau AB &II & 0.37 $\pm$ 0.04 & 0.52 $\pm$ 0.05    &  0.27 $\pm$ 0.03       & S1, field includes  V807 Tau  \\
GI Tau\tablenotemark{f}   & II  & 0.67 $\pm$ 0.07    & 0.50 $\pm$ 0.05    &  0.33 $\pm$ 0.06 &    \\
GK Tau\tablenotemark{f}   & II  & 0.90 $\pm$ 0.09    & 0.59 $\pm$ 0.06    & 0.35 $\pm$ 0.06 &    \\
Haro 6-37 B\tablenotemark{e} &II  & 0.96 $\pm$ 0.10    & 0.90 $\pm$ 0.09    & 1.25 $\pm$ 0.12 &   \\
HBC 347    & III &  $<$ 0.009            &   $<$0.009     &   $<$ 0.019     &   S2 \\
HBC 352/353  &III &  $<$ 0.007      & \nodata           &   $<$ 0.017   &    S1 \\
HBC 354/355 &III  &  $<$ 0.007      & \nodata           &   $<$ 0.022   &    Empty field \\
HBC 356/357&III  &  $<$ 0.010      &  $<$ 0.011      &  $<$ 0.023 &     S1 \\
HBC 358/359 & III &  $<$ 0.007      & \nodata            &  $<$ 0.020 &      S2 \\
HBC 360/361& III &  $<$ 0.009      &  $<$ 0.010  &     $<$ 0.020 &       Empty field \\
HBC 362   & III &   $<$ 0.008       & \nodata            &  $<$ 0.019 &      S4 \\
HBC 372   &III &   $<$ 0.009        &  $<$0.009          &  $<$ 0.021 &      Empty field \\
HBC 376  & III  &   $<$ 0.009        &  $<$ 0.009          &  $<$ 0.016 &        S3 \\
HBC 388   &III &   $<$ 0.007        &  \nodata            &  $<$ 0.018 &   Empty field \\
HBC 392   & III &   $<$ 0.007        & \nodata           &  $<$ 0.013 & Empty field  \\
HBC 407 & III  &  $<$ 0.009 &  $<$ 0.009 &  $<$ 0.017 & Empty field \\
HBC 412 AB & III &  $<$ 0.009 &  $<$ 0.009 &  $<$ 0.018 & Empty field \\
HD 283572 & III & $<$ 0.008 &  $<$ 0.009 &  $<$ 0.027 & S1 \\
HD 283759\tablenotemark{g}& S   & 0.52 $\pm$ 0.05   &   0.69 $\pm$  0.07       & 0.57 $\pm$ 0.06 &   S1 \\
HL Tau  &  I &   75.3 $\pm$ 7.5 & 72.3 $\pm$ 7.2 & 54.7 $\pm$ 8.2 & Cloud emission, 4 T Tauri stars, S1 \\
HN Tau A  &II & 1.02 $\pm$ 0.10 & \nodata & 0.61 $\pm$ 0.06  & S4  \\ 
HO Tau   & II & 0.10 $\pm$ 0.01  & 0.11 $\pm$ 0.01   &  0.14 $\pm$ 0.02  &  S2 \\
HV Tau C\tablenotemark{d}  & I?  & 1.55 $\pm$ 0.16   &  1.68 $\pm$ 0.18\phantom{0}    & 1.45 $\pm$ 0.24 &   \\ 
IQ Tau &  II &  0.74 $\pm$ 0.07 & 0.78 $\pm$ 0.08  &  0.81 $\pm$ 0.08 & S2 \\ 
IW Tau &  III  &  $<$ 0.009          &   $<$ 0.010        &  $<$ 0.030      &  S1 \\
J04305171$+$2441475&  II & 2.85 $\pm$ 0.29 & 3.11 $\pm$ 0.31 & 3.06 $\pm$ 0.31 & Cloud emission, 35\arcsec\ north of ZZ Tau\\
J1-4872 & III &$<$ 0.009          &   $<$ 0.009        &  $<$ 0.027      &  S3 \\
J1-507  &  III &$<$ 0.008    &    $<$ 0.008 &  $<$ 0.021   & Empty field \\
J1-665 & III &  $<$ 0.009          &   $<$ 0.009        &  $<$ 0.025      &  Empty field\\
J2-2041 &III  &$<$ 0.008          &   $<$ 0.009        &  $<$ 0.013      &  S1 \\
JH 108   &  III& $<$ 0.008          &   $<$ 0.009        &  $<$ 0.021      &  S2 \\
JH 223  & II &   0.114 $\pm$ 0.011 &  0.109 $\pm$ 0.011  &  0.076 $\pm$ 0.016 &  \\ 
JH 56     &II &  0.027 $\pm$ 0.003  & 0.017 $\pm$ 0.003  &  $<$ 0.04 &      Cloud emission \\
L 1551-51 &III &  $<$ 0.009          &   $<$ 0.009        &  $<$ 0.016      &  Empty field \\
L 1551-55 &III &  $<$ 0.009          &   $<$ 0.009        &  $<$ 0.016      &  Empty field \\
LkCa 1    & III &  $<$ 0.009        &  $<$ 0.009       &  $<$ 0.021 &    Empty field \\
LkCa 3    & III & $<$ 0.009        &  $<$  0.009       &   $<$ 0.017 &       S1 \\
LkCa 4    & III &  $<$ 0.009        &  $<$ 0.009        &   $<$ 0.034 &   Empty field \\
LkCa 5    & III &  $<$ 0.009        &  $<$ 0.009       &   $<$  0.029 &      S1, Cloud emission \\
LkCa 14   &  III & $<$ 0.009      &   $<$ 0.009          &  $<$ 0.028  &    S2 \\
LkCa 15   & T & 1.23 $\pm$ 0.12    &  1.48 $\pm$ 0.15   & 1.78 $\pm$ 0.18   & \\
LkCa 19   & III &  $<$ 0.009        &  $<$ 0.009        &   $<$ 0.015  &    \\
LkCa 21  &  III & $<$ 0.012 &  \nodata & $<$ 0.051 & in same field as RY Tau \\
RW Aur A  & II & 2.47 $\pm$ 0.15  & 2.94 $\pm$ 0.29 &  1.74 $\pm$ 0.17 &   S2 \\ 
RY Tau   & T? & 14.13 $\pm$ 1.40 & \nodata & 8.81 $\pm$ 0.88 & Cloud emission \\
SAO 76411  & III &  $<$ 0.012          &  $<$ 0.011          &  $<$ 0.020 mJy      S1 \\
SAO 76428 & III  & $<$ 0.009          &  $<$ 0.009          &  $<$ 0.020 mJy      S2 \\
StHA 34\tablenotemark{h} &  II  &  0.053 $\pm$ 0.006 & 0.039 $\pm$ 0.005  & 0.025 $\pm$ 0.025  &  S2 \\
Tau~L1495~14 & I  &1.11 $\pm$ 0.22 & \nodata & 1.45 $\pm$ 0.29 & in same field as V773 Tau, edge of field\\
T Tau NS\tablenotemark{i}  & II/I & 170 $\pm$ 20      & 143  $\pm$   21            & 100 $\pm$ 20  &   extended, cloud emission \\
UZ Tau   Eab+Wab & II &2.03 $\pm$ 0.20    & 2.13 $\pm$ 0.21    & 2.04 $\pm$ 0.20  &   \\
V397 Aur   &  III & 0.016 $\pm$ 0.003  & \nodata           & 0.016 $\pm$ 0.007  &  \\ 
V773 Tau\tablenotemark{b} &  II   & 0.81 $\pm$ 0.08   & \nodata           & 0.25 $\pm$ 0.03 &   \\
V807 Tau AB & II  & 0.51 $\pm$ 0.05 & 0.72 $\pm$ 0.07 & 0.35 $\pm$ 0.04 & S2, field includes GH Tau \\
V819 Tau & II/III & 0.030 $\pm$ 0.003 & 0.017 $\pm$ 0.003  & $<$ 0.050 & S1, faint cloud emission \\ 
V826 Tau  AB & III & $<$ 0.009        &  $<$ 0.009       &  $<$ 0.017 &    Empty field \\
V827 Tau & III & $<$ 0.009        &  $<$ 0.009       &  $<$ 0.018 &    Empty field \\
V830 Tau  & III &  $< $0.009         &   $<$ 0.010          &  $<$ 0.030    &   Cloud emission \\
V836 Tau & T & 0.21 $\pm$ 0.02    & 0.30 $\pm$ 0.03    & 0.27 $\pm$ 0.05  &  S1 \\ 
V927 Tau &III  &   $<$ 0.009        &  $<$ 0.009       &  $<$ 0.032 &    Empty field \\
V928 Tau & III &   $<$ 0.009        &  $<$ 0.011       &  $<$ 0.053 &   Cloud emission \\
V1096 Tau   & III      &   $<$ 0.009     &   $<$ 0.010 &  $<$ 0.030 &  Cloud emission. \\
V1213 Tau & I & 0.41 $\pm$ 0.04 & 0.51 $\pm$ 0.05 & 0.59 $\pm$ 0.012 & in same field as HL Tau\\
VY Tau AB  &  II  & 0.30 $\pm$0.03    & 0.29 $\pm$ 0.03   & 0.22 $\pm$ 0.02 & S2 \\
Wa Tau1   & III & $<$ 0.009          &  $<$ 0.009          &  $<$ 0.014 &       S2  \\
XZ Tau AB & II  & 5.77 $\pm$ 0.87   & 4.71 $\pm$ 0.71     &  4.13 $\pm$ 0.80  & 23\arcsec\ East of HL Tau \\
ZZ Tau  & II & 0.052 $\pm$ 0.005  & 0.051 $\pm$ 0.005    &  0.025 $\pm$ 0.02   & \\ 
\enddata
\tablecomments{Upper limits are 3$\sigma$. The commonly used source notations CIDA NN and J1-NN are not know to SIMBAD, which instead uses [BCG93] NN and [HJS91] NN, respectively. In the
comment field  Sn (n=1,2, 3 etc) lists the number of
``background'' sources we find in each field.}
\tablenotetext{a}{Star surrounded by extended nebulosity, see text.}
\tablenotetext{b}{V773~Tau,  FM~Tau, CW~Tau \& Tau L1495 14 in the same field.}
\tablenotetext{c}{CX~Tau and FP~Tau in the same field.} 
\tablenotetext{d}{DO~Tau and  HV~Tau in the same field. Cloud emission, see text.}
\tablenotetext{e}{Haro~6-37 and DQ~Tau in the same field.}
\tablenotetext{f}{GI~Tau and  GK~Tau in the same field; some nebulosity to the east in red. }
\tablenotetext{g}{Emission is extended! Not a member of the Taurus association \citep{Massarotti05}.}
\tablenotetext{h}{Stronger source within 18\arcsec\ to the NW.}
\tablenotetext{i}{Extended}
\end{deluxetable}

\newpage

\begin{deluxetable}{lcccccc}
\tabletypesize{\scriptsize}
\tabletypesize{\tiny}
\tablecaption{PACS [\ion{O}{1}] 63, [\ion{O}{1}] 145, and [\ion{C}{2}] 158 Flux densities and other possible detections \label{tbl-lineflux}}
\tablewidth{0pt}
\tablehead{
\colhead{Name} &  \colhead{Jet/} & \colhead{SED} &\colhead{[\ion{O}{1}] 63}  & \colhead{[\ion{O}{1}] 145}  & \colhead{[\ion{C}{2}] 158} & \colhead{Other lines}\\
\colhead{}  & \colhead{outflow}&  \colhead{Class} & \colhead{($10^{-16}$ W m$^{-2}$)} & \colhead{($10^{-16}$ W m$^{-2}$)} & \colhead{($10^{-16}$ W m$^{-2}$)} &\colhead{}}
\startdata
AA Tau	& Y& II	&  0.22 $\pm$ 0.02  							& 0.02 $\pm$ 0.01    ($<0.02$)				& $<0.02$								& 1,2,4,5,9\\
BP Tau	&N	& II	&  0.10 $\pm$ 0.03  							& 	\nodata							& 	\nodata							&1\\
CI Tau	&N	& II	&  0.33 $\pm$ 0.05  							& 0.02 $\pm$ 0.02    ($<0.07$)				& $<0.07$								& 9\\
CIDA 2	&N	& III	&  $<0.08$								& 	\nodata							& 	\nodata							&\\
CoKu Tau/4&N	& II	&  0.22 $\pm$ 0.02  							& 	\nodata							& 	\nodata							&\\
CW Tau	&Y	& II	&  0.72 $\pm$ 0.04  							& 0.05 $\pm$ 0.02						& 0.05 $\pm$ 0.01    ($<0.05$)				& 9\\
CX Tau	&N	& T	&  0.07 $\pm$ 0.03  	($<0.08$)    			         & 	\nodata							& 	\nodata							&\\
CY Tau	&N	& II	&  0.12 $\pm$ 0.04 							& 	\nodata							& 	\nodata							&\\
DE Tau	& N	& II	&  0.07 $\pm$ 0.06    ($<0.12$)  				& 0.04 $\pm$ 0.02						& $<0.04$								& 9\\
DF Tau	&Y	& II	&  0.61 $\pm$ 0.06  							& 0.06 $\pm$ 0.04    ($<0.09$)				& 0.05 $\pm$ 0.02    ($<0.06$)				&\\
DG Tau	& Y	& II	&  13.40 $\pm$ 0.17							& 0.84 $\pm$ 0.05						& 1.87 $\pm$ 0.04						& 4,5,7,8, 9\\
DG Tau B	& Y	& I	&  4.26 $\pm$ 0.12  							& 0.27 $\pm$ 0.03						& 0.13 $\pm$ 0.02						& 9\\
DH Tau	&N & II	&  0.07 $\pm$ 0.02    ($<0.10$)			  		& 	\nodata							& 	\nodata							&\\
DI Tau	&N	& III	&  $<0.14$								& 	\nodata							& 	\nodata							&\\
DK Tau	&N	& II	&  0.16 $\pm$ 0.03  							& $<0.10$								& $<0.15$								&\\
DL Tau	& Y	& II	&  0.22 $\pm$ 0.02  							& $<0.14$								& $<0.09$								& 1\\
DM Tau	&N	& T	&  0.07 $\pm$ 0.02  							& $<0.06$								& $<0.06$								& 1\\
DN Tau	&N	& T	&  0.06 $\pm$ 0.02  							& $<0.06$								& 0.05 $\pm$ 0.02    ($<0.09$)				&\\
DO Tau	& Y	& II	&  0.71 $\pm$ 0.10  							& 0.08 $\pm$ 0.01						& 0.03 $\pm$ 0.01    ($<0.03$)				& 2,9\\
DP Tau	& Y	& II	&  1.48 $\pm$ 0.13  							& 0.07 $\pm$ 0.02						& 0.12 $\pm$ 0.02						& 5,9\\
DQ Tau	& N	& II	&  0.21 $\pm$ 0.04  							& $<0.04$								& 0.03 $\pm$ 0.01    ($<0.04$)				&\\
DS Tau	&N	& II	&  0.09 $\pm$ 0.02  							& 	\nodata							& 	\nodata							&\\
FF Tau	&N	& III	&  $<0.10$								& 	\nodata							& 	\nodata							&\\
FM Tau	&N	& II	&  0.10 $\pm$ 0.02    ($<0.13$)				  	& $<0.06$								& $<0.06$								&\\
FO Tau	& N	& T	&  0.12 $\pm$ 0.05    ($<0.12$)				  	& $<0.05$								& $<0.04$								&\\
FQ Tau	&N	& II	&  $<0.09$	  							& $<0.03$								& $<0.06$								&\\
FS Tau A	& Y	& II/Flat	&  3.58 $\pm$ 0.05  						& 0.18 $\pm$ 0.01						& 0.16 $\pm$ 0.01						& 1,4,5,7,8,9,10,11\\
FT Tau	&N	& II	&  0.17 $\pm$ 0.05  							& 	\nodata							& 	\nodata							&\\
FW Tau	&N	& III	&  0.04 $\pm$ 0.02    ($<0.08$)				  	& 	\nodata							& 	\nodata							&\\
FX Tau	&N	& II 	&  $<0.14$						  		& 	\nodata							& 	\nodata							&\\
GG Tau Aab	&N	& II	&  0.51 $\pm$ 0.04  							& 0.03 $\pm$ 0.01    ($<0.05$)				& $<0.04$								& 4,5,9\\
GH Tau	&N	& II	&  0.08 $\pm$ 0.04    ($<0.16$)				 	& $<0.03$								& $<0.02$								&\\
GI/GK Tau\tablenotemark{a}	&N{\bf /Y}	& II	&  0.31 $\pm$ 0.14  							& 	\nodata							& 	\nodata							&\\
GM Aur	&N	& T	&  0.24 $\pm$ 0.05  							& $<0.06$								& $<0.10$								&\\
GO Tau	&N	& II	&  $<0.06$							  	& 	\nodata							& 	\nodata							&\\
Haro 6-5B	&Y	& I	&  0.68 $\pm$ 0.11  							& 0.04 $\pm$ 0.01    ($<0.04$)				& 0.06 $\pm$ 0.01    						&\\
Haro 6-13	&Y	& II	&  0.70 $\pm$ 0.05  							& 0.03 $\pm$ 0.01    ($<0.04$)				& $<0.05$								& 2,4,5,9\\
Haro 6-37	 &N	& II	&  0.10 $\pm$ 0.02  							& 	\nodata							& 	\nodata							&\\
HBC 347&N	         & III		&  $<0.12$								& 	\nodata							& 	\nodata							&\\
HBC 356	&N	& III	&  $<0.08$								& 	\nodata							& 	\nodata							&\\
HBC 358	&N	& III	&  $<0.14$								& 	\nodata							& 	\nodata							&\\
HD 283572&N	& III	&  $<0.08$								& 	\nodata							& 	\nodata							&\\
HK Tau	&N	& II	&  0.34 $\pm$ 0.02  							& $<0.04$								& $<0.04$								& 9\\
HL Tau	&Y	& I	&  5.13 $\pm$ 0.05  							& 0.28 $\pm$ 0.02						& 0.36 $\pm$ 0.04						& 1,2,3,4,5,7,8,9\\
HN Tau A	& Y	& II	&  0.41 $\pm$ 0.02  							& 0.04 $\pm$ 0.01						& $<0.04$								& 1\\
HO Tau	&N	& II	&  $<0.10$								& 	\nodata							& 	\nodata							&\\
HV Tau C	&Y	& I?	&  0.52 $\pm$ 0.08  							& 	\nodata							& 	\nodata							&\\
IP Tau	&N	& T	&  0.06 $\pm$ 0.02    ($<0.07$)  				& 	\nodata							& 	\nodata							&\\
IQ Tau	&N	& II	&  0.15 $\pm$ 0.03  							& $<0.11$								& $<0.07$								&\\
IRAS 04158+2805&Y& 	I &  0.57 $\pm$ 0.04  						& 0.05 $\pm$ 0.02						& $<0.08$							 	&\\
IRAS 04385+2550 & Y& II	&  0.49 $\pm$ 0.02  						& 	\nodata							& 	\nodata							&\\
J1-4827    &N	& III	&  $<0.09$								& 	\nodata							& 	\nodata							&\\
LkCa 1	&N	& III	&  $<0.08$								& 	\nodata							& 	\nodata							&\\
LkCa 3	&N	& III	&  $<0.10$								& 	\nodata							& 	\nodata							&\\
LkCa 4	&N	& III	&  $<0.10$								& 	\nodata							& 	\nodata							&\\
LkCa 5	&N	& III	&  $<0.10$								& 	\nodata							& 	\nodata							&\\
LkCa 7	&N	& III	&  $<0.09$								& 	\nodata							& 	\nodata							&\\
LkCa15	&N	& T	&  0.10 $\pm$ 0.02  							& $<0.09$								& $<0.11$								&\\
RW Aur 	&Y	& II	&  1.54 $\pm$ 0.05  							& 0.12 $\pm$ 0.03						& $<0.10$								& 9,10\\
RY Tau	&Y	& T?	&  1.05 $\pm$ 0.05  							& 0.07 $\pm$ 0.02						& $<0.10$								& 1,2,9\\
SU Aur	&Y	& II &  0.86 $\pm$ 0.03  							& 0.06 $\pm$ 0.02    ($<0.06$)				& 0.07 $\pm$ 0.02						& 9\\
T Tau	&Y	& II/I	&  81.80 $\pm$ 0.31							& 4.06 $\pm$ 0.04						& 2.29 $\pm$ 0.02						&1,2,3,4,5,6,7,8,9,10,11\\
UX Tau A&N		& T &  0.34 $\pm$ 0.02  							& 	\nodata							& 	\nodata							&\\
UY Aur	&Y	& II	&  3.14 $\pm$ 0.04  							& 0.17 $\pm$ 0.01						& 0.16 $\pm$ 0.01						&1,2,4,5,7,8,9,10\\
UZ Tau	&Y	& II	&  0.45 $\pm$ 0.14  							& $<0.11$								& $<0.07$								&\\
V710 Tau	&N	& II	&  0.10 $\pm$ 0.06    ($<0.12$)  				& 	\nodata							& 	\nodata							&\\
V773 Tau	&Y	& II	&  0.65 $\pm$ 0.03  							& 	\nodata							& 	\nodata							& 1\\
V807 Tau	& N	& II	&  0.14 $\pm$ 0.05 							& $<0.05$								& 0.02 $\pm$ 0.01 ($<0.04$)				&\\
V819 Tau	&N	&  II/III	&  $<0.09$							& 	\nodata							& 	\nodata							&\\
V836 Tau	&N	& T &  0.05 $\pm$ 0.04    ($<0.06$)				  	& 	\nodata							& 	\nodata							&\\
V927 Tau	&N	& III	&  $<0.08$								& 	\nodata							& 	\nodata							&\\
V1096 Tau	&N & III	&  $<0.12$								& 	\nodata							& 	\nodata							&\\
VY Tau	&N	& II &  $<0.11$								& 	\nodata							& 	\nodata							&\\
XZ Tau	&Y	& II	&  3.61 $\pm$ 0.09							& 0.18 $\pm$ 0.04						& 0.24 $\pm$ 0.02						& 1,2,4,5,7,8,9\\
ZZ Tau	&N	& II &  $<0.10$								& 	\nodata							& 	\nodata							&\\
\enddata
\tablenotetext{a}{Gi Tau and GK Tau are unresolved in spectroscopy.}

\tablecomments{For sources where no line is detected, the 3$\sigma$ upper limit is reported.\\
Possible detections of other emission lines determined through visual inspection. The numbers refer to:\\
1)  o-H$_{2}$O $8_{18}\rightarrow7_{07}$\\
2) CH$^{+}$ J = $5 \to 4$ \\
3) CO J = $36 \to 35$\\
4) o-H$_{2}$O	$4_{23}\rightarrow3_{12}$\\ 
5) OH 1/2-3/2 hfs \\
6) CO J = $33 \to 32$\\
7) p-H$_{2}$O $3_{22}\rightarrow2_{11}$\\
8) CO J = $29 \to 28$\\
9) CO J = $18 \to 17$\\ 
10) o-H$_{2}$O $2_{12}\rightarrow1_{01}$\\
11) o-H$_{2}$O $2_{21}\rightarrow2_{12}$}

\end{deluxetable}

\newpage
\begin{deluxetable}{lcccclrl}
\tabletypesize{\scriptsize}
\tablecaption{PACS Sectroscopy Settings \label{tbl-spec}}
\tablewidth{0pt}
\tablehead{
\colhead{Obs. Mode}	&	\colhead{Setting}	&	\colhead{Grating order}	&	\colhead{Camera}	&	\colhead{Observed Range}	&	\colhead{Species}	&	\colhead{Transition}	&	\colhead{Wavelength}\\
\colhead{} 			&	\colhead{} 		& 	\colhead{} 			& 	\colhead{} 		& 	\colhead{($\mu$m)}			& 	\colhead{}  		& 	\colhead{} 		&	\colhead{($\mu$m)} }
\startdata
LineSpec					&	A 	&	3	&	Blue	&	$62.68-63.68$			&	{\bf [OI]}			&	${\bf ^{3}P_{1}\rightarrow^{3}P_{2}}$		&	{\bf 63.184}	\\
						&		&		&		&						&	o-H$_{2}$O		&	$8_{18}\rightarrow7_{07}$				&	63.324		\\
						&		&	1	&	Red	&	$188.77-190.30$		&	DCO$^{+}$		&	J = $22 \to 21$								&	189.570		\\
\hline
RangeSpec				&	B	&	2	&	Blue	&	$72.00-73.05$			&	CH$^{+}$			&	J =$5 \to 4	$							&	72.14		\\
						&		&		&		&						&	CO				&	J = $36 \to 35$								&	72.843		\\
						&		&	1	&	Red	&	$144-146.1$			&	CO				&	J = $18 \to 17$								&	144.784		\\			
						&		&		&		&						&	{\bf [OI]}			&	${\bf ^{3}P_{0}\rightarrow^{3}P_{1}}$		&	{\bf 145.525}	\\
\hline			
						&	C	&	2	&	Blue	&	$78.55-79.45$			&	o-H$_{2}$O		&	$4_{23}\rightarrow3_{12}$				&	78.741		\\
						&		&		&		&						&	OH				&	$\frac{1}{2} \to \frac{3}{2}$ hfs							&	79.11/79.18	\\
						&		&		&		&						&	CO				&	J = $33 \to 32$								&	79.360		\\
						&		&	1	&	Red	&	$157.1-158.9$			&	{\bf [CII]}			&	${\bf ^{2}P_{3/2}\rightarrow^{2}P_{1/2}}$		&	{\bf 157.741}	\\
						&		&		&		&						&	p-H$_{2}$O		&	$3_{31}\rightarrow4_{04}$				&	158.309		\\	
\hline			
						&	D	&	2	&	Blue	&	$89.45-90.50$			&	p-H$_{2}$O		&	$3_{22}\rightarrow2_{11}$				&	89.988		\\
						&		&		&		&						&	CH+				&	J = $4 \to 3$								&	90.02		\\
						&		&		&		&						&	CO				&	J = $29 \to 28$								&	90.163		\\
						&		&	1	&	Red	&	$178.9-181.0$			&	o-H$_{2}$O		&	$2_{12}\rightarrow1_{01}$				&	179.527		\\
						&		&		&		&						&	CH+				&	J = $2 \to 1$								&	179.610		\\
						&		&		&		&						&	o-H$_{2}$O		&	$2_{21}\rightarrow2_{12}$				&	180.488
\enddata
\end{deluxetable}

\begin{deluxetable}{lccccc}
\tabletypesize{\small}
\tablecaption{GASPS Taurus Observations: PACS Spectroscopy \label{tbl-obs}}
\tablewidth{0pt}
\tablehead{
\colhead{Target} & \colhead{RA (J2000)}& \colhead{Dec (J2000)}& \colhead{Obs. Mode} & \colhead{Exposure}& \colhead{ObsID}\\
\colhead{} &	\colhead{(h m s)}& \colhead{(deg m s)} & \colhead{} & \colhead{(s)} &\colhead{} }
\startdata
AA Tau 	                            & 4 34 55.420 	& +24 28 53.16 & LineSpec 	& 1252 	& 1342190357 \\
					& 	 		&			& LineSpec	& 6628	& 1342225758 \\
					&	 		&			& RangeSpec 	& 5141	& 1342190356 \\
	 				&      			&			& RangeSpec 	&20555	& 1342225759\\
BP Tau 				& 4 19 15.830 	& +29 06 26.90 & LineSpec 	& 1252 	& 1342192796 \\
					&			&			& LineSpec 	& 3316	& 1342225728	\\
CIDA 2                                  & 4 15 05.157	& +28 08 46.21	&  LineSpec  	& 1252     &  1342216643\\					
CI Tau 				& 4 33 52.000 	& +22 50 30.20 & LineSpec 	& 1252	& 1342192125 \\
					&			&			& RangeSpec 	& 5141	& 1342192124 \\
CoKu Tau/4 			& 4 41 16.800 	& +28 40 00.60 & LineSpec 	& 1252 	& 1342191360 \\  
					&			&			& LineSpec	& 6628	& 1342225837 \\
CW Tau				& 4 14 17.000 	& +28 10 57.80 & LineSpec 	& 1252 	& 1342216221 \\
					&			&			& RangeSpec 	& 10279	& 1342216222 \\
CX Tau				& 4 14 47.860 	& +26 48 11.01	& Linespec	& 3316	& 1342225729	\\
CY Tau 				& 4 17 33.730 	& +28 20 46.90 & LineSpec 	& 1252 	& 1342192794 \\
DE Tau 				& 4 21 55.640 	& +27 55 06.10 & LineSpec 	& 1252 	& 1342192797 \\
					&			&			& RangeSpec 	& 8316 	& 1342216648\\
DF Tau 				& 4 27 03.080 	& +25 42 23.30 & LineSpec 	& 1252 	& 1342190359 \\
					&			&			& RangeSpec 	& 5141	& 1342190358 \\
DG Tau 				& 4 27 04.700 	& +26 06 16.30 & LineSpec 	& 1252	& 1342190382 \\
					&			&			& RangeSpec 	& 5141	& 1342190383 \\
DG Tau B 				& 4 27 02.560 	& +26 05 30.70 & LineSpec 	& 1252 	& 1342192798 \\
					&			&			& RangeSpec 	& 10279	& 1342196652 \\
DH Tau				& 4 29 42.020 	& +26 32 53.20 & LineSpec	& 1252	& 1342225734 \\
DI Tau\tablenotemark{d}	& 4 29 42.475	& +26 32 49.31	 & LineSpec	& 1252	& 1342225734 \\
DK Tau 				& 4 30 44.240 	& +26 01 24.80 & LineSpec 	& 1252 	& 1342192132 \\
					&			&			& LineSpec 	& 3316	& 1342225732	\\
					&			&			& RangeSpec 	& 5141	& 1342192133 \\
DL Tau  		& 4 33 39.060 	& +25 20 38.23 & LineSpec 	& 1252 	& 1342190355 \\
					&			&			& LineSpec	& 6628	& 1342225800 \\
					&			&			& RangeSpec 	& 5141	& 1342190354 \\
DM Tau 				& 4 33 48.720 	& +18 10 10.00 & LineSpec 	& 1252 	& 1342192123 \\
					&			&			& LineSpec	& 6628	& 1342225825 \\
					&			&			& RangeSpec 	& 5141	& 1342192122 \\
DN Tau 				& 4 35 27.370 	& +24 14 58.90 & LineSpec 	& 1252 	& 1342192127 \\
					&			&			& LineSpec	& 3316	& 1342225757 \\
					&			&			& RangeSpec 	& 5141	& 1342192126 \\
DO Tau 				& 4 38 28.580 	& +26 10 49.44 & LineSpec 	& 1252 	& 1342190385 \\
					&			&			& RangeSpec 	& 5141	& 1342190384 \\
					&			&			& RangeSpec 	&12516	& 1342225802\\
DP Tau				& 4 42 37.700 	& +25 15 37.50 & LineSpec 	& 1252 	& 1342191362 \\
					&			&			& RangeSpec 	& 10279	& 1342225827\\
DQ Tau				& 4 46 53.050	& +17 00 00.20	& LineSpec	& 1252	& 1342225806 \\
					&			&			& RangeSpec 	& 8316	& 1342225807\\
DS Tau				& 4 47 48.110	& +29 25 14.45	& LineSpec	& 3316	& 1342225851\\
FF Tau 				& 4 35 20.900	& +22 54 24.20 &LineSpec 	& 1252 	& 1342192802 \\
FM Tau 				& 4 14 13.580 	& +28 12 49.20 & LineSpec 	& 1252 	& 1342216218 \\
					&			&			& RangeSpec 	& 10279 	& 1342216219\\
FO Tau 				& 4 14 49.290 	& +28 12 30.60 & LineSpec 	& 1252 	& 1342216645 \\
					&			&			& RangeSpec 	& 8316 	& 1342216644\\
FQ Tau 				& 4 19 12.810 	& +28 29 33.10 & LineSpec 	& 1252 	& 1342192795 \\
					&			&			& RangeSpec 	& 8316	& 1342216650\\
FS Tau				& 4 22 02.180 	& +26 57 30.50 & LineSpec 	& 1252 	& 1342192791 \\
					&			&			& RangeSpec 	& 10279	& 1342194358 \\
FT Tau 				& 4 23 39.190 	& +24 56 14.10 & LineSpec 	& 1252 	& 1342192790 \\
FW Tau				& 4 29 29.710	& +26 16 53.20	& LineSpec	& 1252	& 1342225735 \\
FX Tau 				& 4 30 29.610 	& +24 26 45.00 &LineSpec 	& 1252 	& 1342192800 \\
GG Tau 				& 4 32 30.350 	& +17 31 40.60 & LineSpec 	& 1252 	& 1342192121 \\
					&			&			& RangeSpec 	& 5141	& 1342192120 \\
					&			&			& RangeSpec 	& 12516	& 1342225738\\
GH Tau 				& 4 33 06.430 	& +24 09 44.50 & LineSpec 	& 1252 	& 1342192801 \\
					&			&			& RangeSpec 	& 8316	& 1342225762\\
GI Tau/GK Tau			& 4 33 34.310	& +24 21 11.50	& LineSpec	& 3316	& 1342225760	\\
GM Aur 				& 4 55 10.990 	& +30 21 59.25 & LineSpec 	&1252 	& 1342191357 \\
					&			&			& RangeSpec 	& 5141	& 1342191356 \\
GO Tau 				& 4 43 03.090 	& +25 20 18.80 & LineSpec 	& 1252 	& 1342191361 \\
					&			&			& LineSpec	& 3316	& 1342225826 \\
Haro 6-5B\tablenotemark{a}			& 4 22 02.18 	& +26 57 30.5	& LineSpec	& 1252	& 1342192791 \\
Haro 6-13			& 4 32 15.410 	& +24 28 59.70 & LineSpec 	& 1252 	& 1342192128 \\
					&			&			& RangeSpec 	& 5141	& 1342192129 \\
					&			&			& RangeSpec 	& 12516	& 1342225761\\
Haro 6-37				& 4 46 58.980	& +17 02 38.20	& LineSpec	& 1252	& 1342225805 \\
HBC 347 				& 3 29 38.370 	& +24 30 38.00 & LineSpec 	& 1252 	& 1342192136 \\
HBC 356 				& 4 03 13.990 	& +25 52 59.90 &LineSpec 	& 1252 	& 1342204134 \\
  		 			& 			& 			& LineSpec 	& 1252 	& 1342214359 \\
HBC 358 				& 4 03 50.840 	& +26 10 53.20 & LineSpec 	& 1252 	& 1342204347 \\
  		 			& 			& 			& LineSpec 	& 1252 	& 1342214680 \\
HD 283572                          & 4 21 58.847  & +28 18 06.51 & LineSpec       & 1660     &  1342216646\\
HK Tau				& 4 31 50.570	& +24 24 18.07	& LineSpec 	& 3316	& 1342225736 \\
					&			&			& RangeSpec 	& 8316	& 1342225737\\
HL Tau\tablenotemark{b}	& 4 31 38.437	& 18 13 57.65	& LineSpec	& 1252	& 1342190351	\\
HN Tau			& 4 33 39.350	& +17 51 52.37	& LineSpec	& 3316	& 1342225796 \\
					&			&			& RangeSpec 	& 10279	& 1342225797\\
HO Tau 				& 4 35 20.200 	& +22 32 14.60 & LineSpec 	& 1252 	& 1342192803 \\
HV Tau				& 4 38 35.280	& +26 10 38.63	& LineSpec	& 3316	& 1342225801 \\
IP Tau				& 4 24 57.080	& +27 11 56.50	& LineSpec	& 3316	& 1342225756 \\
IQ Tau 				& 4 29 51.560 	& +26 06 44.90 & LineSpec 	& 1252 	& 1342192135 \\
					&			&			& LineSpec	& 3316	& 1342225733 \\
					&			&			& RangeSpec 	& 5141	& 1342192134 \\
IRAS 04158+2805		& 4 18 58.140 	& +28 12 23.50 & LineSpec 	& 1252 	& 1342192793 \\
					&			&			& RangeSpec 	& 5141	& 1342192792 \\
IRAS 04385+2550		& 4 41 38.820	& +25 56 26.75	& LineSpec	& 3316	& 1342225828 \\
					&			&			& RangeSpec 	& 8316	& 1342225829\\
J1- 4872                        & 4 25 17.678	& +26 17 50.41 & LineSpec       & 1660     &  1342216653\\
LkCa 1                                  & 4 13 14.142	& +28 19 10.84& LineSpec  	& 1660     &  1342214679 \\
LkCa 3                                  & 4 14 47.973	& +27 52 34.65	& LineSpec   	 & 1660     &  1342216220 \\
LkCa 4                                  & 4 16 28.109	& +28 07 35.81	& LineSpec       & 1660     &  1342216642 \\
LkCa 5                                  & 4 17 38.940	& +28 33 00.51	& LineSpec       & 1660     &  1342216641 \\
LkCa 7                                  & 4 19 41.272	& +27 49 48.49	&  LineSpec      & 1660     &  1342216649\\
LkCa 15 				& 4 39 17.800 	& +22 21 03.48 & LineSpec 	& 1252 	& 1342190387 \\
					&			&			& LineSpec	& 6628	& 1342225798 \\
					&			&			& RangeSpec 	& 5141	& 1342190386 \\
RW Aur			& 5 07 49.540 	& +30 24 05.07 & LineSpec 	&1252 	& 1342191359 \\	
					&			&			& RangeSpec 	& 5141	& 1342191358 \\
RY Tau				& 4 21 57.400 	& +28 26 35.54 & LineSpec 	& 1252 	& 1342190361 \\
					&			&			& RangeSpec 	& 5141	& 1342190360 \\
SU Aur				& 4 55 59.380 	& +30 34 01.56 & LineSpec 	& 3316 	& 1342217844 \\
					&			&			& RangeSpec 	&10279	& 1342197845 \\
T Tau				& 4 21 59.430 	& +19 32 06.37 & LineSpec 	&1252 	& 1342190353 	\\
					&			&			& RangeSpec 	& 5141	& 1342190352 \\
					&			&			&LineSpec 	& 1252	& 1342215699 \\
UX Tau 				& 4 30 03.990 	& +18 13 49.40 & LineSpec 	& 1252 	& 1342204350 \\
    	 				& 	 		& 			 & LineSpec 	& 1252 	& 1342214357 \\
UY Aur				& 4 51 47.380 	& +30 47 13.50 & LineSpec 	& 1252 	& 1342193206 \\
					&			&			& LineSpec 	& 1252	& 1342215699 \\
					&			&			& RangeSpec 	& 10279 	& 1342226001\\ 	
UZ Tau				& 4 32 42.890 	& +25 52 32.60 & LineSpec 	& 1252 	& 1342192131 \\
					&			&			& RangeSpec 	& 5141	& 1342192130 \\
V710 Tau 				& 4 31 57.800 	& +18 21 35.10 & LineSpec 	& 1252 	& 1342192804 \\
V773 Tau			& 4 14 12.920 	& +28 12 12.45 & LineSpec 	& 3316 	& 1342216217 \\
V807 Tau\tablenotemark{c}& 4 33 06.641 & +24 09 54.99 & LineSpec 	& 1252 	& 1342192801 \\
					&			&			& RangeSpec 	& 8316	& 1342225762\\
V819 Tau 				& 4 19 26.260 	& +28 26 14.30 & LineSpec 	& 1660 	& 1342216651 \\
V836 Tau                             & 5 03 06.600  &  +25 23 19.70 & LineSpec     & 3316      &  1342227634\\
V927 Tau				& 4 31 23.820	& +24 10 52.93	& LineSpec	& 3316	& 1342225763\\
V1096 Tau                           & 4 13 27.230	& +28 16 24.80	& LineSpec	& 1252     &  1342214678\\
VY Tau 				& 4 39 17.410 	& +22 47 53.40 & LineSpec 	& 1252 	& 1342192989 \\
XZ Tau  			& 4 31 39.480	& +18 13 55.70 & LineSpec 	& 1252 	& 1342190351 \\
					&			&			& RangeSpec 	& 5141	& 1342190350 \\
ZZ Tau 				& 4 30 51.380 	& +24 42 22.30 & LineSpec 	& 1252 	& 1342192799 \\
\enddata

\tablenotetext{a}{Haro 6-5B observed in the same field of view as FS Tau}
\tablenotetext{b}{HL Tau observed in the same field of view as XZ Tau}
\tablenotetext{c}{V807 Tau observed in the same field of view as GH Tau}
\tablenotetext{d}{DI Tau observed in the same field of view as DH Tau}
\end{deluxetable}

\begin{deluxetable}{lcccccccc}
\tabletypesize{\scriptsize}
\tablecaption{Continuum Flux in spectroscopic observations\label{tbl-contflux}}
\label{ContFlux}
\tablewidth{0pt}
\tablehead{
\colhead{Name} & \colhead{63.18 $\mu$m}  & \colhead{72.84 $\mu$m}  & \colhead{78.74 $\mu$m} & \colhead{90.16 $\mu$m} & \colhead{145.53 $\mu$m} & \colhead{157.74 $\mu$m} & \colhead{179.53 $\mu$m} & \colhead{189.57 $\mu$m} \\
\colhead{} & \colhead{[Jy]} & \colhead{[Jy]} & \colhead{[Jy]} & \colhead{[Jy]} & \colhead{[Jy]} & \colhead{[Jy]} & \colhead{[Jy]} & \colhead{[Jy]} }

\startdata
AA Tau			& 0.85$\pm$0.02		& 0.85$\pm$0.02	& 0.82$\pm$0.02	& 0.89$\pm$0.01	& 0.99$\pm$0.01	& 1.18$\pm$0.01	& 1.12$\pm$0.02	& 0.92$\pm$0.05	\\
BP Tau			& 0.45$\pm$0.04		&	\nodata		&	\nodata		&	\nodata		&	\nodata		&	\nodata		&	\nodata		& $<$0.17      		\\
CI Tau 			& 1.38$\pm$0.05		& 1.62$\pm$0.03  	& 1.57$\pm$0.05	& 1.90$\pm$0.04 	& 2.04$\pm$0.03 	& 2.37$\pm$0.03 	& 2.52$\pm$0.04	& 2.06$\pm$0.07	\\
CIDA 2			& $<$0.09				&	\nodata		&	\nodata		&	\nodata		&	\nodata		&	\nodata		&	\nodata		& $<$0.24			\\
CoKu Tau-4	 	& 0.90$\pm$0.02		&	\nodata		& 	\nodata		& 	\nodata		& 	\nodata		& 	\nodata		& 	\nodata		& 0.61$\pm$0.04	\\
CW Tau			& 1.31$\pm$0.04		& 1.48$\pm$0.03	& 1.58$\pm$0.02	& 1.59$\pm$0.03	& 1.67$\pm$0.02	& 1.93$\pm$0.02	& 1.83$\pm$0.02	& 1.81$\pm$0.10	\\
CX Tau			& 0.30$\pm$0.03		&	\nodata		& 	\nodata		& 	\nodata		& 	\nodata		& 	\nodata		& 	\nodata		& $<$0.22     	 	\\
CY Tau		 	& $<$0.15				&	\nodata		& 	\nodata		& 	\nodata		& 	\nodata		& 	\nodata		& 	\nodata		& $<$0.29			\\
DE Tau			& 1.35$\pm$0.05		& 0.70$\pm$0.03	& 0.67$\pm$0.02	&	\nodata		& 0.65$\pm$0.01	& 0.62$\pm$0.01	&	\nodata		& $<$0.27			\\
DF Tau		 	& 0.43$\pm$0.05		& 0.32$\pm$0.03	& 0.35$\pm$0.04	& 0.59$\pm$0.04	& 0.20$\pm$0.02	& 0.12$\pm$ 0.03	& 0.20$\pm$0.04	& $<$0.29			\\
DG Tau 		 	& 13.94$\pm$0.09		& 11.94$\pm$0.04	& 13.72$\pm$0.07	& 14.87$\pm$0.04	& 14.09$\pm$0.03	& 15.27$\pm$0.03	& 14.26$\pm$0.05	& 11.54$\pm$0.06	\\
DG Tau B			& 9.85$\pm$0.05		& 10.37$\pm$0.03	& 10.96$\pm$0.04	& 11.79$\pm$0.03	& 14.63$\pm$0.02	& 15.34$\pm$0.02	& 14.46$\pm$0.03	& 13.88$\pm$0.09	\\
DH Tau			& 0.25$\pm$0.05		&	\nodata		& 	\nodata		& 	\nodata		& 	\nodata		& 	\nodata		& 	\nodata		&  0.39$\pm$0.07	\\
DI Tau			& $<$0.15				&	\nodata		&	\nodata		&	\nodata		&	\nodata		&	\nodata		&	\nodata		& $<$0.22			\\
DK Tau			& 0.86$\pm$0.03		& 0.73$\pm$0.04	& 0.73$\pm$0.04	& 0.90$\pm$0.05	& 0.76$\pm$0.03	& 0.85$\pm$0.03	& 1.04$\pm$0.04	& $<$0.31 		\\
DL Tau			& 0.96$\pm$0.02		& 0.71$\pm$0.03	& 1.07$\pm$0.05	& 1.08$\pm$0.05	& 1.49$\pm$0.02	& 1.77$\pm$0.03	& 2.15$\pm$0.04	& 1.58$\pm$0.03	\\
DM Tau			& 0.58$\pm$0.02		& 0.68$\pm$0.03	& 0.86$\pm$0.04	& 0.81$\pm$0.03	& 0.81$\pm$0.03	& 0.93$\pm$0.02	& 1.04$\pm$0.04	& 0.65$\pm$0.03	\\
DN Tau			& 0.69$\pm$0.02		& 0.55$\pm$0.05	& 0.70$\pm$0.05	& 0.58$\pm$0.04	& 0.63$\pm$0.02	& 0.77$\pm$0.03	& 0.78$\pm$0.04	& 0.70$\pm$0.06	\\
DO Tau		 	& 2.82$\pm$0.05		& 4.01$\pm$0.02	& 4.16$\pm$0.02	& 3.15$\pm$0.04	& 3.94$\pm$0.01	& 4.27$\pm$0.01	& 3.45$\pm$0.04	& 2.87$\pm$0.07	\\
DP Tau		 	& 0.79$\pm$0.06		& 0.36$\pm$0.02	& 0.32$\pm$0.03	& 0.37$\pm$0.02	& 0.18$\pm$0.02	& 0.25$\pm$0.02	& 0.32$\pm$0.03	& $<$0.31		 	\\
DQ Tau			& 1.14$\pm$0.04		& 1.13$\pm$0.02	& 1.18$\pm$0.02	& 	\nodata		& 0.98$\pm$0.01 	& 1.09$\pm$0.02	& 	\nodata		&  0.64$\pm$0.09	\\
DS Tau			& 0.17$\pm$0.03		&	\nodata		& 	\nodata		& 	\nodata		& 	\nodata		& 	\nodata		& 	\nodata		& $<$0.17			\\
FF Tau			& $<$0.12				&	\nodata		&	\nodata		&	\nodata		&	\nodata		&	\nodata		&	\nodata		& $<$0.29			\\
FM Tau			& 0.56$\pm$0.06		& 0.39$\pm$0.02	& 0.39$\pm$0.02	& 0.59$\pm$0.03	& 0.23$\pm$0.02	& 0.30$\pm$0.02	& 0.26$\pm$0.02	& $<$0.29			\\
FO Tau			& 0.42$\pm$0.05		& 0.35$\pm$0.02	& 0.39$\pm$0.02	& 	\nodata		& 0.23$\pm$0.01	& 0.26$\pm$0.01	& 	\nodata		&  $<$0.23		\\
FQ Tau			& $<$0.12		  		&  0.11$\pm$0.02	& 0.15$\pm$0.02	& 	\nodata		& 0.06$\pm$0.01	& 0.09$\pm$0.01	& 	\nodata		& $<$0.32 		\\
FS Tau A		 	& 2.34$\pm$0.05		& 1.73$\pm$0.03	& 1.61$\pm$0.07	& 1.88$\pm$0.03	& 1.63$\pm$0.02	& 1.78$\pm$0.02	& 1.55$\pm$0.03	& 1.47$\pm$0.09	\\
FT Tau		 	& 0.52$\pm$0.04		&	\nodata		& 	\nodata		& 	\nodata		& 	\nodata		& 	\nodata		& 	\nodata		& 1.06$\pm$0.11	\\
FW Tau			& $<$0.12				&	\nodata		& 	\nodata		& 	\nodata		& 	\nodata		& 	\nodata		& 	\nodata		& $<$0.29			\\
FX Tau			& 0.17$\pm$0.05		&	\nodata		& 	\nodata		& 	\nodata		& 	\nodata		& 	\nodata		& 	\nodata		& $<$0.29			\\
GG Tau Aab	& 2.93$\pm$0.04		& 3.97$\pm$0.01	& 4.51$\pm$0.03	& 4.52$\pm$0.04	& 7.62$\pm$0.01	& 8.60$\pm$0.01	& 8.22$\pm$0.03	& 7.13$\pm$0.08	\\
GH Tau			& 0.87$\pm$0.05		&  0.34$\pm$0.02	& 0.30$\pm$0.02 	& 	\nodata		& 0.18$\pm$0.02	& 0.20$\pm$0.01	& 	\nodata		& 0.31$\pm$0.07	\\
GI/GK Tau\tablenotemark{a}			& 0.91$\pm$0.07		&	\nodata		& 	\nodata		& 	\nodata		& 	\nodata		& 	\nodata		& 	\nodata		&  0.82$\pm$0.12	\\
GM Aur		 	& 1.99$\pm$0.05		& 2.46$\pm$0.05  	& 2.46$\pm$0.04  	& 3.00$\pm$0.04  	& 3.64$\pm$0.02  	& 4.25$\pm$0.03  	& 5.00$\pm$0.04	& 2.78$\pm$0.07	\\
GO Tau			& 0.24$\pm$0.02 		&	\nodata		&	\nodata		&	\nodata		&	\nodata		&	\nodata		&	\nodata		& 0.50$\pm$0.04	\\
Haro6-5B			& 2.11$\pm$0.05		& 1.87$\pm$0.02	& 1.68$\pm$0.03	& 1.79$\pm$0.03	& 2.26$\pm$0.02	& 2.61$\pm$0.02	& 2.66$\pm$0.03	& 3.73$\pm$0.10	\\
Haro 6-13		 	& 5.04$\pm$0.05		& 5.53$\pm$0.05 	& 6.59$\pm$0.02  	& 5.84$\pm$0.04   	& 6.37$\pm$0.01  	& 6.86$\pm$0.02  	& 6.55$\pm$0.05	& 5.41$\pm$0.08	\\
Haro 6-37			& 0.85$\pm$0.04		&	\nodata		& 	\nodata		& 	\nodata		& 	\nodata		& 	\nodata		& 	\nodata		& 1.42$\pm$0.07	\\
HBC 347			& $<$0.15				&	\nodata		&	\nodata		&	\nodata		&	\nodata		&	\nodata		&	\nodata		& $<$0.32			\\
HBC 356			& $<$0.09				&	\nodata		& 	\nodata		& 	\nodata		& 	\nodata		& 	\nodata		& 	\nodata		& $<$0.24			\\
HBC 358			& $<$0.18				&	\nodata		& 	\nodata		& 	\nodata		& 	\nodata		& 	\nodata		& 	\nodata		& $<$0.29			\\
HD 283572		& $<$0.09				&	\nodata		&	\nodata		&	\nodata		&	\nodata		&	\nodata		&	\nodata		& $<$0.22			\\
HK Tau			& 2.03$\pm$0.03		& 2.13$\pm$0.02	& 2.29$\pm$0.02	& 	\nodata		& 2.13$\pm$0.01	& 2.43$\pm$0.01 	& 	\nodata		& 1.73$\pm$0.07	\\
HL Tau		 	& 66.19$\pm$0.14		& 65.15$\pm$0.07	& 65.98$\pm$0.08    	& 67.79$\pm$0.08    	& 65.91$\pm$ 0.06 	& 69.37 $\pm$0.08 	& 53.90$\pm$0.09	& 45.45$\pm$0.11	\\
HN Tau			& 0.77$\pm$0.02		& 0.71$\pm$0.02	& 0.79$\pm$0.03	& 0.74$\pm$0.03	&  0.60$\pm$0.01	& 0.64$\pm$0.02	& 0.60$\pm$0.03	& 0.55$\pm$0.07 	\\
HO Tau			& $<$0.12				&	\nodata		&	\nodata		&	\nodata		&	\nodata		&	\nodata		&	\nodata		& $<$0.34			\\
{\bf HV Tau C}		& 0.80$\pm$0.04		&	\nodata		&	\nodata		&	\nodata		&	\nodata		&	\nodata		&	\nodata		& 1.09$\pm$0.05 	\\
IP Tau			& 0.44$\pm$0.03		&	\nodata		&	\nodata		&	\nodata		&	\nodata		&	\nodata		&	\nodata		& $<$0.16			\\
IQ Tau			& 0.61$\pm$0.03		& 0.48$\pm$0.06	& 0.43$\pm$0.03	& 0.59$\pm$0.04	& 0.59$\pm$0.02	& 0.70$\pm$0.03	& 0.76$\pm$0.04	& 0.71$\pm$0.06	\\
IRAS 04158+2805	& 0.76$\pm$0.05		& 0.92$\pm$0.03	& 0.93$\pm$0.04	& 1.25$\pm$0.05	& 2.45$\pm$0.02	& 3.18$\pm$0.03	& 3.87$\pm$0.03	& 2.94$\pm$0.08	\\
{\bf IRAS 04358+2550}	& 1.87$\pm$0.02		&	\nodata		&	\nodata		& 	\nodata		&	\nodata 		&	\nodata 		&	\nodata 		& 1.44$\pm$0.04	\\
{\bf J1-4827}		& $<$0.09				&	\nodata		&	\nodata		&	\nodata		&	\nodata		&	\nodata		&	\nodata		& $<$0.27			\\
LkCa 1			& $<$0.09				&	\nodata		&	\nodata		& 	\nodata		&	\nodata 		&	\nodata 		&	\nodata 		& $<$0.26			\\
LkCa 3			& $<$0.12				&	\nodata		&	\nodata		& 	\nodata		&	\nodata 		&	\nodata 		&	\nodata 		& $<$0.27			\\
LkCa 4			& 0.12$\pm$0.04		&	\nodata		&	\nodata		& 	\nodata		&	\nodata 		&	\nodata 		&	\nodata 		& $<$0.22			\\ 
LkCa 5			& $<$0.12				&	\nodata		&	\nodata		& 	\nodata		&	\nodata 		&	\nodata 		&	\nodata 		& $<$0.24			\\
LkCa 7			& $<$0.12				&	\nodata		&	\nodata		& 	\nodata		&	\nodata 		&	\nodata 		&	\nodata 		& $<$0.20			\\
LkCa 15		 	& 0.99$\pm$0.02		& 0.98$\pm$0.04	& 1.08$\pm$0.04	& 1.13$\pm$0.04	& 1.55$\pm$0.03	& 1.82$\pm$0.03	& 2.18$\pm$0.06	& 1.28$\pm$0.04	\\
RW Aur		 	& 1.79$\pm$0.05		& 1.85$\pm$0.04  	& 2.09$\pm$0.04  	& 2.04$\pm$0.04  	& 1.30$\pm$0.02   	& 1.54$\pm$0.03  	& 1.41$\pm$0.04	& 1.12$\pm$0.13	\\
RY Tau		 	& 10.86$\pm$0.07		& 9.82$\pm$0.03	& 10.10$\pm$0.04	& 10.00$\pm$0.04	& 7.98$\pm$0.02	& 8.64$\pm$0.03	& 8.50$\pm$0.04	& 5.73$\pm$0.11	\\
SU Aur			& 5.95$\pm$0.02		& 5.44$\pm$0.02	& 4.72$\pm$0.03	& 4.11$\pm$0.03	& 3.24$\pm$0.02	& 3.33$\pm$0.02	& 2.10$\pm$0.02	& 1.49$\pm$0.05	\\
T Tau		 	& 86.12$\pm$0.21		& 77.48$\pm$0.07	& 75.93$\pm$0.10	& 67.18$\pm$0.15	& 39.53$\pm$0.26	& 39.64$\pm$0.06	& 35.77$\pm$0.23	& 25.28$\pm$0.13 	\\
UX Tau		 	& 3.28$\pm$0.03		&	\nodata		& 	\nodata		& 	\nodata		& 	\nodata		& 	\nodata		& 	\nodata		& 2.31$\pm$0.07	\\
UY Aur		 	& 4.85$\pm$0.06		& 4.87$\pm$0.02	& 4.89$\pm$0.04	& 4.52$\pm$0.04	& 3.11$\pm$0.01	& 3.30$\pm$0.02	& 2.39$\pm$0.02	& 2.11$\pm$0.07	\\
UZ Tau			& 1.23$\pm$0.06		& 1.29$\pm$0.04 	& 1.47$\pm$0.05	& 1.38$\pm$0.04	& 1.78$\pm$0.03	& 1.94$\pm$0.02	& 1.69$\pm$0.04	& 1.54$\pm$0.09	\\
V710 Tau			& 0.47$\pm$0.05		&	\nodata		&	\nodata		&	\nodata		&	\nodata		&	\nodata		&	\nodata		& 1.10$\pm$0.18	\\
V773 Tau			& 0.66$\pm$0.03		&	\nodata		&	\nodata		&	\nodata		&	\nodata		&	\nodata		&	\nodata		& $<$0.17			\\
V807 Tau			& 0.58$\pm$0.05		& 0.13$\pm$0.02	&  0.12$\pm$0.02	&	\nodata		& 0.17$\pm$0.02	& 0.18$\pm$0.02	&	\nodata		& $<$0.29			\\
V819 Tau      		& $<$0.12				&	\nodata		&	\nodata		&	\nodata		&	\nodata		&	\nodata		&	\nodata		&$<$0.22			\\
V836 Tau			& 0.33$\pm$0.02		&	\nodata		&	\nodata		&	\nodata		&	\nodata		&	\nodata		&	\nodata		& 0.27$\pm$0.04	\\
V927 Tau			& $<$0.09				&	\nodata		&	\nodata		&	\nodata		&	\nodata		&	\nodata		&	\nodata		& $<$0.17			\\
V1096 Tau		& $<$0.15				&	\nodata		&	\nodata		&	\nodata		&	\nodata		&	\nodata		&	\nodata		& $<$0.27			\\
VY Tau			& $<$0.12				&	\nodata		&	\nodata		&	\nodata		&	\nodata		&	\nodata		&	\nodata		& $<$0.29 		\\					
XZ Tau			& 4.20$\pm$0.05		& 3.98$\pm$0.04	& 4.04$\pm$0.06	& 3.89$\pm$0.04	& 2.90$\pm$0.02	& 2.78$\pm$0.03	& 2.87$\pm$0.06	& 3.71$\pm$0.08	\\
ZZ Tau			& $<$0.12				&	\nodata		&	\nodata		&	\nodata		&	\nodata		&	\nodata		&	\nodata		& $<$0.29	\\		
\enddata
\tablenotetext{a}{Gi Tau and GK Tau are unresolved in spectroscopy.}

\end{deluxetable}

\end{document}